# An Automated Framework for the Extraction of Semantic Legal Metadata from Legal Texts

**Amin Sleimi@** · **Nicolas Sannier** ·
**Mehrdad Sabetzadeh** · **Lionel Briand** ·
**Marcello Ceci** · **John Dann**

**Abstract** Semantic legal metadata provides information that helps with understanding and interpreting legal provisions. Such metadata is therefore important for the systematic analysis of legal requirements. However, manually enhancing a large legal corpus with semantic metadata is prohibitively expensive. Our work is motivated by two observations: (1) the existing requirements engineering (RE) literature does not provide a harmonized view on the semantic metadata types that are useful for legal requirements analysis; (2) automated support for the extraction of semantic legal metadata is scarce, and it does not exploit the full potential of artificial intelligence technologies, notably natural language processing (NLP) and machine learning (ML). Our objective is to take steps toward overcoming these limitations. To do so, we review and reconcile the semantic legal metadata types proposed in the RE literature. Subsequently, we devise an automated extraction approach for the identified metadata types using NLP and ML. We evaluate our approach through two case studies over the Luxembourgish legislation. Our results indicate a high accuracy in the generation of metadata annotations. In particular, in the two case studies, we were able to obtain precision scores of 97.2% and 82.4%, and recall scores of 94.9% and 92.4%.

**Keywords** Legal Requirements · Semantic Legal Metadata · Natural Language Processing (NLP)

## 1 Introduction

Legal metadata provides explicit conceptual knowledge about the content of legal texts. The requirements engineering (RE) community has long been interested in legal metadata as a way to systematize the process of identifying and elaborating legal compliance requirements (Breaux and Antón, 2008; Massey et al., 2010; Zeni et al., 2015). The relevant literature identified several facets of legal metadata: *administrative metadata* keeps track of the

Amin Sleimi, Nicolas Sannier, Marcello Ceci
SnT, University of Luxembourg
Mehrdad Sabetzadeh, Lionel Briand
School of EECS, University of Ottawa & SnT, University of Luxembourg
John Dann
Central Legislative Service (SCL), Government of Luxembourg.

lifecycle of a legal text, e.g., the text's creation date, its authors, its effective date, and its history of amendments; *provenance metadata* maintains information about the origins of a legal text, e.g., the parliamentary discussions preceding the ratification of a legislative text; *usage metadata* links legal provisions to their applications in case law, jurisprudence, and doctrine; *structural metadata* captures the hierarchical organization of a legal text (or legal corpus); finally, *semantic metadata* captures fine-grained information about the meaning and interpretation of legal provisions, e.g., modalities (obligation, prohibition and permission), actors, conditions, exceptions, and violations.

Among the above, structural and semantic metadata have received the most attention in RE. Structural metadata is used mainly for establishing traceability to legal provisions, and for performing tasks such as requirements change impact analysis (Gordon and Breaux, 2012; Sannier et al., 2017) and prioritization (Massey et al., 2010; Massey, 2012). Semantic metadata is a prerequisite for the systematic derivation of compliance requirements (Breaux and Antón, 2008; Breaux, 2009; Siena et al., 2009; Bhatia et al., 2016a) and for transitioning from legal texts to formal specifications (Maxwell and Antón, 2010) or models (Zeni et al., 2015; Siena et al., 2009; Zeni et al., 2016).

In this article, we concern ourselves with *semantic legal metadata*. In Fig. 1, we exemplify such metadata over three illustrative legal statements. These statements come from the traffic laws for Luxembourg, and have been translated into English from their original language, French. Statement 1 concerns the management of public roads by the municipalities. Statement 2 concerns penalties for violating the inspection processes for vehicles. Statement 3 concerns the interactions between the magistrates in relation to ongoing prosecutions on traffic offenses. In these examples, we provide metadata annotations only for the phrases within the statements (*phrase-level metadata*). Some of these phrase-level annotations, however, can also induce annotations at the level of statements (*statement-level metadata*): for example, the "may" modality in Statements 1 and 3 makes these statements permissions, and the modal verb "shall" in Statement 2, combined with the presence of a sanction, makes the statement a penalty statement. The metadata types illustrated in Fig. 1 will be further explained in Section 3.

The example statements in Fig. 1 entail legal requirements for various governmental IT systems, including road and critical infrastructure management systems, as well as case processing applications used by the law enforcement agencies and the courts. In this regard, the metadata annotations in Fig. 1 provide useful information to requirements analysts: as we argue more precisely in Section 2, the RE literature identifies several use cases for semantic legal metadata in the elicitation and elaboration of legal requirements. For instance, the annotations of Statement 1 may help with finding the conditions under which a road restriction can be put in place. The annotations of Statement 2 may lead the analyst to define a compliance rule composed of an antecedent (i.e., absence of an agreement), an action (i.e., performing vehicle inspections) and a consequence (i.e., a range of sanctions). Finally, the annotations of Statement 3 provide cues about the stakeholders who may need to be interviewed during requirements elicitation (agents and auxiliary parties), as well as the way in which these stakeholders should interact, potentially using IT systems.

Our work in this article is motivated by two observed limitations in the state-of-the-art on semantic legal metadata, discussed below.

*1) Lack of a harmonized view of semantic legal metadata for RE.* While the RE community acknowledges the importance of semantic legal metadata, there is no consensus on the metadata types that are beneficial for legal requirements analysis. Different work strands propose different metadata types (Breaux, 2009; Zeni et al., 2015; Maxwell and Antón,

1. Within the limits and according to the provisions stated in this article, the municipal authorities may, in whole or in part, temporarily or permanently, regulate or prohibit traffic on the public roads of the territory of the municipality, provided that these municipal regulations concern the traffic on the municipal roads as well as on the national roads situated inside the municipality's agglomerations.

2. One who performs vehicle inspections without being in possession of the agreement specified in paragraph 1 shall be punished with an imprisonment of eight days to three years and a fine of 251 to 25,000€ or one of these penalties only.

3. The investigating judge may pronounce the prohibition of driving at the request of the public prosecutor against a person sued for an offense under this Act or for an offense or a crime associated with one or more contraventions of the traffic regulations on any public road.

Fig. 1: Examples of Semantic Legal Metadata Annotations

2010; Kiyavitskaya et al., 2008; Siena et al., 2012), but no strand completely covers the others.

*2) Lack of coverage of recent advances in NLP and ML.* If done manually, enhancing a large corpus of legal texts with semantic metadata is extremely laborious. Recently, increasing effort has been put into automating this task using natural language processing (NLP). Notable initiatives aimed at providing automation for metadata extraction are GaiusT (Zeni et al., 2015) and NomosT (Zeni et al., 2016). These initiatives do not handle the broader set of metadata types proposed in the RE literature, e.g., locations (Breaux, 2009), objects (Massey et al., 2010), and situations (Siena et al., 2012). Besides, they rely primarily on simple NLP techniques, e.g., tokenization, named-entity recognition, and part-of-speech (POS) tagging. Although these techniques have the advantage of being less prone to mistakes, they cannot provide detailed insights into the complex semantics of legal provisions.

With recent developments in NLP, the robustness of advanced NLP techniques, notably constituency and dependency parsing, has considerably improved (Hirschberg and Manning, 2015). This raises the prospect that these more advanced techniques may now be accurate enough for a deep automated analysis of legal texts. Dependency parsing is important for correctly identifying constituents whose roles are influenced by linguistic dependencies. For instance, in Statement 3 of Fig. 1, the roles of the "(sued) person", the "investigating judge" and the "public prosecutor" can be derived from such dependencies. Constituency parsing is instead important for delineating phrases out of simpler nouns or chunks in a statement. For instance, in Statement 1 of Fig. 1, annotating "the national roads situated inside the municipality's agglomerations" as one segment requires the ability to recognize this segment as a compound noun phrase. Without a parse tree, one cannot readily mark this segment in its entirety.

In addition, machine learning (ML) provides a potentially useful mechanism for distinguishing metadata types that NLP-based rules cannot handle with sufficient accuracy. For instance, in Statement 3 of Fig. 1, the "investigating judge", the "public prosecutor" and the "sued person" hold three closely related stakeholder roles, differentiated by whether they are acting, are a third party or are the target of the action described in the statement. Articulating explicit rules for distinguishing such metadata types proved very difficult. This prompted us to investigate whether ML can be employed for telling apart such metadata types.

In this article, we take a step toward addressing the two limitations outlined above by developing a framework for automated semantic legal metadata extraction. We start by reviewing and reconciling several existing metadata classifications in order to devise a conceptual model of semantic metadata types pertinent to legal requirements analysis. This model defines six metadata types for legal statements and 18 metadata types for the phrases contained therein. Next, we perform a qualitative study over 200 legal statements from the traffic laws of Luxembourg in order to define rules that can automatically detect the metadata contained in legal statements. This qualitative study results in a set of NLP-based rules for automated extraction of semantic legal metadata, covering the majority of the phrase-level metadata types and the statement-level metadata types of the conceptual model. These rules are complemented by using ML classification techniques in order to further distinguish stakeholders' roles in the statements. In our evaluation, we analyze 350 differents statements from six different legislative domains in order to assess the accuracy of our metadata extraction rules.

This article is an extension of a previous conference paper (Sleimi et al., 2018) published at the 26th IEEE International Requirements Engineering Conference (RE 2018). The main extensions in this article are:

1. A detailed description of the conceptual model of semantic metadata types pertinent to legal requirements analysis, presented in our previous work. More precisely, we detail the reconciliation work that we have performed over the concepts retrieved from the literature in order to build our conceptual model, and present a glossary of the metadata types in our conceptual model.
2. The extension of our initial set of extraction rules to increase the coverage of our conceptual model.
3. An extended description of the technical aspects of our approach.
4. A deeper evaluation of our initial case study concerning the extraction of semantic metadata from the traffic laws in Luxembourg.
5. New empirical evidence for the accuracy of our approach through a second case study including 200 additional statements randomly selected from five other legislative domains (commerce, environment, health, labour, and penal codes). This investigation is a stepping stone toward assessing the generalizability of our approach.

*Overview and Structure.* Fig. 2 summarizes our approach. In the figure, the activities and artifacts in white were already presented in our previous work (Sleimi et al., 2018), whereas those in grey are novel or extended. Section 2 presents the literature on the elicitation of relevant legal concepts. Section 3 discusses the mapping of the concepts from the literature to each other in order to reconcile the corresponding semantics, and presents a conceptual model of legal concepts based on the mapping. Section 4 introduces a qualitative study aimed at validating the metadata types of the conceptual model by observing their occurrence in the legal statements. An additional objective of this qualitative study is to determine cues that could help in automating the extraction of the metadata types. Section 5 addresses the elaboration of extraction rules for the legal concepts in our conceptual model using a combination of NLP and ML. Section 6 describes the implementation of the extraction rules into a tool for semantic legal metadata extraction. Section 7 empirically assesses, through two case studies, the accuracy of our extraction rules. Section 8 discusses threats to validity, and Section 9 concludes the article.

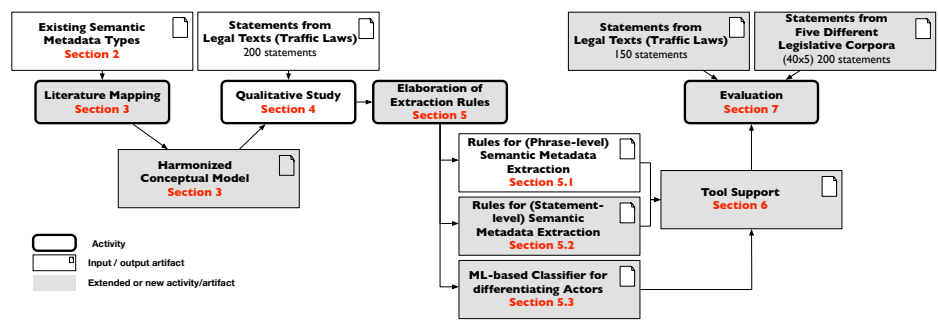

Fig. 2: Approach Overview

## 2 Background and Related Work

We begin with background information on deontic logic and the Hohfeldian system. These serve as the foundations for most work in the area of legal analysis. Subsequently, we discuss the related work on semantic legal metadata. Finally, we position our technical approach by explaining how constituency and dependency parsing have been used previously in RE.

### 2.1 Preliminaries

When trying to interpret and analyze the semantics of the law, most existing research is rooted in either deontic logic (Horty, 2001) or in the Hohfeldian system of legal concepts (Hohfeld, 1917). Deontic logic distinguishes "what is permitted" (permission) from "what ought to be done" (obligation) and what is "not permitted" (prohibition).

The Hohfeldian system (Hohfeld, 1917) distinguishes eight terms for legal positions: right, privilege, power, immunity, duty, no-right, liability, and disability. Each term in the Hohfeldian system is paired with one opposite and one correlative term. Hohfeldian opposites contradict each other. For example, the opposite of privilege is duty: the permission to park in a given zone is the opposite of the prohibition to park in the same zone. Hohfeldian correlatives instead entail each other. For example, the right of a party entails that there is another party (a counter-party) who has a correlative duty: a driver has the right to know why their vehicle has been stopped by the police, which entails a correlative duty for the police to explain the reason for stopping the vehicle.

## 2.2 Semantic Metadata in Legal Requirements

Deontic logic and the Hohfeldian system introduce a number of important legal concepts. Several strands of work leverage these concepts for the elicitation and specification of legal requirements, and for the definition of compliance rules. Below, we outline these strands and their underlying legal concepts. Examples for many of the legal concepts can be found in Fig. 1. However, we note that not all publications provide precise definitions for the concepts they use. Furthermore, the provided definitions for certain concepts vary across different publications. Consequently, while Fig. 1 is useful for illustrating existing work, the definitions used by others may not be fully aligned with ours. Our definitions for the concepts in Fig. 1 are based on the conceptual model that we propose in Section 3.

*Early foundations.* Two of the earliest research strands in RE on extracting information from legal texts (Giorgini et al., 2005; Breaux et al., 2006) target the elicitation of rights and permissions following the principles of deontic logic. The work by Breaux et al. provides also a proof-of-concept of how structured information may be extracted from legal texts. Extending the generic **Cerno** information extraction framework (Kiyavitskaya et al., 2006), another work by the same authors (Kiyavitskaya et al., 2008) develops automation for the approach of Breaux et al.; the automation addresses *rights*, *obligations*, *exceptions*, *constraints*, *cross-references*, *actors*, *policies*, *events*, *dates*, and *information*.

The above strands lay the groundwork for two different branches of research on legal requirements. The first branch is oriented around goal modeling, and the second around formal rules specified in either restricted natural language or logical form.

*Goal-based legal requirements.* The initial work of Kiyavitskaya et al. with Cerno was enhanced by the **GaiusT tool** (Zeni et al., 2015). GaiusT pursues the explicit objective of identifying metadata in legal texts and using this metadata for building goal-based representations of legal requirements. GaiusT is centered around the concepts of: (1) *actors* who have *goals*, *responsibilities* and *capabilities*; (2) *prescribed behaviors* according to the concepts of *right*, *obligation*, and their respective opposites; (3) *resources*, specialized into *assets* and *information*; (4) *actions* that describe what is taking place; (5) *constraints*, either *exceptions* or *temporal conditions*, which affect the *actors*, *prescribed behaviors* or *resources*. GaiusT further addresses structural legal metadata, which are not covered here.

In tandem with GaiusT, the different versions of the **Nomos framework** (Siena et al., 2009; Zeni et al., 2016; Siena et al., 2012; Ingolfo et al., 2014) provide a complementary perspective toward metadata extraction with a more pronounced alignment with goal models. Nomos models are built around five core concepts: *role* (the holder or beneficiary of a provision), *norm* (either *duty* or *right*), *situation* (describing the past, actual or future state of the world), and *association* (describing how a provision affects a given situation). Zeni et al. propose **NomosT** (Zeni et al., 2016) to automate the extraction of Nomos concepts using GaiusT. While still grounded in Nomos' original concepts, NomosT reuses several other concepts from GaiusT, including *actor*, *resource*, *condition*, and *exception*.

The strands of work presented above follow the principles of deontic logic. Another strand of work on goal-based analysis of legal requirements is **LegalGRL** (Ghanavati et al., 2014; Ghanavati, 2013) which, in contrast to the above, follows the Hohfeldian system. The main legal concepts in LegalGRL are: *subject*, *modality* (based on Hohfeld's classifications of legal positions), *verb*, *action*, *cross-reference*, *precondition*, and *exception*. LegalGRL does not yet have automated support for metadata extraction.

*Formal legal requirements.* Following up on their earlier work (Breaux et al., 2006) and motivated by deriving compliance requirements, Breaux et al. propose an **upper ontology** for formalizing "frames" in legal provisions (Breaux, 2009; Breaux and Antón, 2008). This ontology has two tiers. The first tier describes statement-level (sentence-level) concepts: *permission*, *obligation*, *refrainment*, *exclusion*, *fact*, and *definition*. The second tier describes the concepts related to the constituent phrases in legal statements (phrase-level concepts). In this second tier, *actions* are used as containers for the following concepts: *subject*, *act*, *object*, *purpose*, *instrument* and *location*. For actions that are *transactions*, one or more *targets* need to be specified. Breaux et al. further consider *modalities*, *conditions* and *exceptions* at the level of phrases.

Maxwell and Antón propose a classification of semantic concepts for building formal representations of legal provisions (Maxwell and Antón, 2010). These representations are meant to guide analysts through requirements elicitation. At the level of statements, the classification envisages the concepts of *right*, *permission obligation* and *definition*. At phrase level, the concepts of interest are the *actors* involved in a provision and the *preconditions* that apply to the provision.

Massey et al. develop an approach for mapping the terminology of a legal text onto that of a requirements specification (Massey, 2012; Massey et al., 2010). The goal here is to assess how well legal concerns are addressed within a requirements specification. Massey et al. reuse the concepts of *right*, *obligation*, *refrainment* and *definition* from Breaux et al.'s upper ontology, while adding *prioritization*. At phrase level, the approach uses *actor*, *data object*, *action* and *cross-reference*.

## 2.3 Semantic Metadata in Legal Knowledge Representation

There is considerable research in the legal knowledge representation community on formalizing legal knowledge (Boella et al., 2016). Several ontologies have been developed for different dimensions of legal concepts (Peters et al., 2007; Sartor et al., 2013). Our goal here is not to give a thorough exposition of these ontologies, because our focus is on the metadata types (discussed in Section 2.2) for which clear use cases exist in the RE community. Nonetheless, an overall understanding of the major initiatives in the legal knowledge representation community is important for our purposes, for two reasons. First, these initiatives serve as a confirmatory measure to ensure that we define our metadata types at the right level of abstraction. Second, by considering these initiatives, we are able to create a mapping between the metadata types used in RE and those used in these initiatives, which is a helpful step toward bridging the gap between the two communities.

We consider two major initiatives, LKIF (Hoekstra et al., 2007; Breuker et al., 2006; Boer et al., 2007) and LegalRuleML (Athan et al., 2013; Lam et al., 2016), which are arguably the largest attempts to date at harmonizing legal concepts.

**LKIF** is a rule-modeling language for a wide spectrum of legal texts, ranging from legislation to court decisions. LKIF's core ontology includes over 200 classes. At statement level, LKIF supports the following deontic concepts: *right*, *permission*, *obligation*, and *prohibition*. At phrase level, LKIF's most pertinent concepts are: *actor*, *object*, *event*, *time*, *location*, *trade*, *transaction*, and *delegation* (further specialized into *mandate* and *assignment*). LKIF further provides concepts for the *antecedent* and *consequent* of events.

**LegalRuleML** (Athan et al., 2013; Lam et al., 2016) – a successor of LKIF – tailors the generic RuleML language (RuleML, 2015) for the legal domain. LegalRuleML classifies statements into *facts* and *norms*. Norms are further specialized into *constitutive statements* (definitions), *prescriptive statements*, and *penalty statements*. The modality of a prescriptive statement is expressed using one of the following concepts: *right*, *permission*, *obligation*, or *prohibition*. Penalty statements have embedded into them the concepts of *violation* and *reparation*. LegalRuleML introduces the following additional concepts directly at phrase level: *participant*, *event*, *time*, *location*, *jurisdiction*, *artifact*, and *compliance* (opposite of *violation*). The participants may be designated as *agents*, *bearers* or *third parties*, who may have *roles* and be part of an *authority*.

All the above-mentioned concepts from LKIF and LegalRuleML have correspondences in the RE literature on legal requirements, reviewed in Section 2.2. In Section 3 we reconcile all the RE-related legal concepts found in the literature, in an attempt to provide a unified model of legal metadata for RE.

2.4 Constituency and Dependency Parsing in RE

As mentioned before, the main NLP techniques that we employ for enabling metadata extraction are constituency and dependency parsing. In recent years, these – and other – advanced NLP techniques have generated a lot of traction in RE. Examples of problems to which these techniques have been applied are: template conformance checking (Arora et al., 2015), model extraction (Lucassen et al., 2017), feature extraction (Quirchmayr et al., 2018), and ambiguity and defect detection (Elrakaiby et al., 2017; Rosadini et al., 2017).

In relation to legal requirements, other strands of work (Bhatia et al., 2016a,b; Evans et al., 2017) apply constituency and dependency parsing for analyzing privacy policies. Even though these approaches provide useful inspiration, our objective is different. These strands of work focus on detecting ambiguities in privacy policies via the construction of domain-specific lexicons and ontologies. Our work, in contrast, addresses the extraction of metadata for facilitating the identification and specification of legal requirements, thus aligning best with the GaiusT and NomosT initiatives discussed earlier. What distinguishes our work from these initiatives is the wider coverage of metadata types and the use of NLP techniques to increase accuracy in identifying the spans for metadata annotations.

3 A Model of Semantic Legal Metadata

In this section, we present the mapping of the concepts elicited in Section 2. This mapping constitutes the basis for the elaboration of our conceptual model for legal concepts.

The main challenge in reconciling the above proposals is that they introduce distinct but overlapping concepts. We present our mapping of legal concepts in Table 1. For the elaboration of the mapping, we consider the concepts elicited in the related work (columns of the table), analyze their definition (when provided) and compare them to the definitions contained in other work, in an attempt to reconcile the concepts and the terminology.

In the table, concepts in bold and shaded green have a definition that is close to our own related concept. For instance, the concepts of *exception*, *right*, *obligation* and *definition* are shared and are similar across most of the examined work.

Concepts in italic and shaded orange are related to the concepts in our own taxonomy, but the alignment between the concepts is weaker than for the ones in bold and shaded green. This is largely due to variations in the granularity level adopted across the examined work. For instance, while one can find notions for actors in most of the work, the granularity at which they are described, i.e., their role, varies from one strand of work to another. It varies from an explicit list of roles in Breaux et al.'s upper ontology (Breaux and Antón, 2008), to the simple notions of role in Nomos (Siena et al., 2012) and of actor in Cerno (Kiyavitskaya et al., 2006) and GaiusT (Zeni et al., 2015), and to the fine-grained taxonomy in LKIF (Hoekstra et al., 2007). Another example are obligations and prohibitions, which in Nomos (Siena et al., 2012) are represented by the single concept of *norm (duty)*.

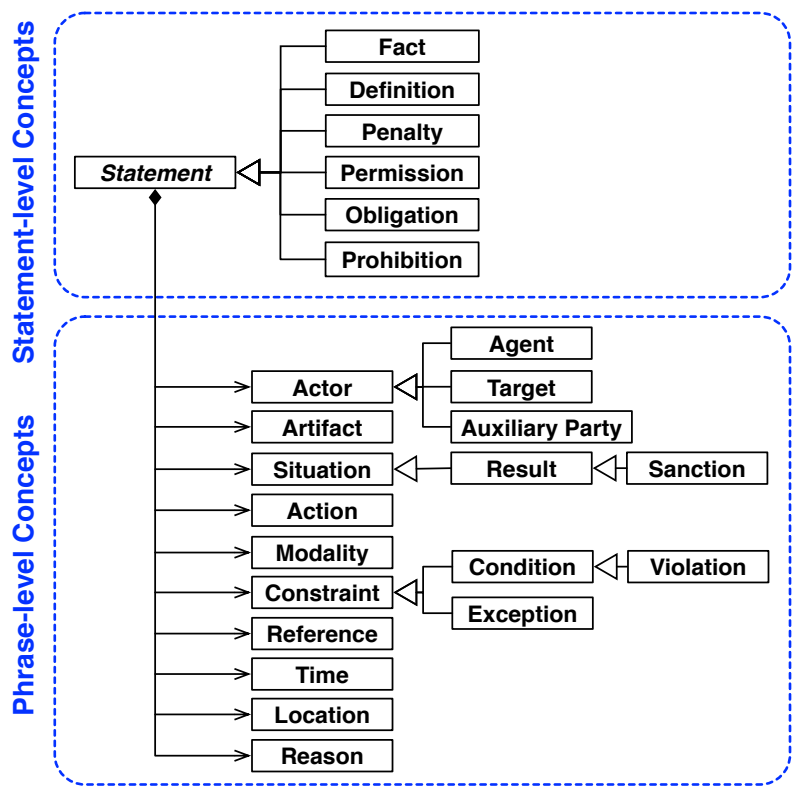

Fig. 3: Conceptual Model for Semantic Legal Metadata Relevant to RE

Empty cells correspond to concepts that are not described in a given work, or whose abstraction is too far from our interpretation. For instance, only few strands of work are concerned with *definition*, *violation*, *penalty* or *sanction*.

In the first column, there are concepts (highlighted in red) that correspond to concepts in the literature that we initially characterized but eventually decided not to retain. The rationale for these decisions is discussed during the elaboration of our conceptual model.

Our conceptual model for semantic legal metadata is presented in Fig. 3. It leverages the mapping that we have performed and described in the previous Section. The dashed boundaries in the figure distinguish statement-level and phrase-level metadata types. Our conceptual model brings together existing proposals from the literature (Breaux and Antón, 2008; Maxwell and Antón, 2010; Siena et al., 2012; Massey et al., 2010; Ghanavati et al., 2014; Zeni et al., 2015). The model derives the vast majority of its concepts (83.3% or 20/24) from Breaux et al.'s upper ontology (Breaux and Antón, 2008) and GaiusT (Zeni et al., 2015).

Our model includes six concrete concepts at statement level. Aside from *penalty*, all statement-level concepts are from Breaux et al.'s upper ontology (Breaux and Antón, 2008).

Table 1: Mapping of the Various Legal Concepts Elicited in Selected Work from the Literature

| Our Approach | LegalRuleML [30, 31] | LKIF [27] | Cerno [20] | Nomos [8, 13, 21] | GaiusT [3] | NomosT [11] | Breaux's Upper Ontology [1, 7] | Massey's Classification [2] | Maxwell's Classification [10] | Legal GRL [22, 23] |
|---|---|---|---|---|---|---|---|---|---|---|
| Agent | Agent | Agent | Actor | Role (Holder) | Agent | Role (Holder) / Actor | Subject | Actor | | Subject (Actor) |
| Artifact | | Artifact | Information | | Resource / Information / Asset | | | Data Object | | |
| | | Natural Object | | | Resource / Information / Asset | | | | | |
| | | Document / evidence | Information | | Resource / Information / Asset | | | Data Object | | |
| | Authority | Public Body / Legislative Body | | | | | | | | |
| Auxiliary Party | Auxiliary Party | Organization / Co-operative | | | | Actor | Target | Actor | Actor | |
| | Auxiliary Party | Person | Actor | Role | Actor | Role | Target | Actor | Actor | |
| Compliance | Compliance | | | Satisfies Norm/situation | | | | | | |
| Constraint / Condition | Context | Cause | Constraint | | Constraint | Condition | Condition | | Precondition | Precondition |
| Definition | Constitutive Statement | | | | | | Definition | Definition | Definition | |
| Delegation | | Assignment / Delegation / Mandate / Trade | | | | | Transaction | | | |
| Exception | | Exception | Exception | Block norm | Exception | Block norm / Derogation | Exception | Exception | | Exception |
| Fact | Factual Statement | Fact | | | | | Fact | | | |
| Location / Jurisdiction | Jurisdiction | Place | | | | | Location | | | |
| Modality | | | | | | | Modality | | | Modality |
| Obligation | Obligation | Obligation | Obligation | Norm (Duty) | Obligation | Norm (Duty) | Obligation | Obligation | Obligation | Duty/Claim – Immunity/Disability* |
| Exclusion | | Immunity | Anti-obligation | | Anti Obligation | | Exclusion | | | Immunity/Disability |
| Penalty | Penalty Statement | | | | | | | | | |
| Permission | Permission | Permission / allowed | | | | | Permission | | Permission | Privilege/No Claim – Power/Liability* |
| Prescriptive provision | Prescriptive Statement | | | | | | | | | Hohfeldian Statement |
| Prohibition | Prohibition | Prohibition | Anti-right | Norm (Right) | Anti Right | Norm (Right) | | | | |
| Reason | | Reason | | | Goal | | Purpose | | | |
| Reference | Legal Source / Reference | Legal Source | Policy / Cross-Reference | | | | | Cross-Reference | Cross-Reference | Cross-Reference |
| Result | | | | Activate norm / Consequent | | Activate norm / Consequent | | Refrainment | | |
| Sanction | Reparation | | | | | | Refrainment | | | |
| Right | Right | Right | Right | Norm (Right) | Right | Norm (Right) | | Right | Right | Power/Liability |
| Role | Role | Role | | Role | Actor | Role / Actor | | Actor | Actor | |
| Action / Situation | | Action / Process | Event | Situation / Antecedent | Action | Situation / Antecedent | Act / Instrument | Action | | Clause (verb)/ Clause (action) |
| Target | Bearer | | | Role (beneficiary) | Actor | Role (Beneficiary) | Object / Target | Actor | Actor | Subject (Exception Actor) |
| Time | Time / Temporal | Time | Date | | Temporal Condition | | | | | |
| Violation | Violation | Disallowed | | Break Norm/situation | | Break Norm/situation | | | | |
| - | Override | | | | | | | Prioritization | | |
| - | | Application / Efficacy | | | | | Quality | | | |

* according to the authors

Penalty comes from LKIF (Hoekstra et al., 2007): we found this concept to be a necessary designation for statements containing sanctions. The model envisages 18 concrete concepts for phrases, most of which are illustrated in the statements of Fig. 1. We propose a definition for each of these concepts in Table 2 and we further discuss them below, starting with statement-level concepts.

*Fact* is something that is known or proved to be true and comes from Breaux et al.'s upper ontology (Breaux and Antón, 2008). This is in line with the classifications from Legal-RuleML (Athan et al., 2013) and LKIF (Hoekstra et al., 2007).

Table 2: Glossary for Our Legal Concepts

| Concept | Definition |
| --- | --- |
| Action | the process of doing something |
| Actor | an entity that has the capability to act |
| Agent | an entity that is the main actor performing the action |
| Artifact | a human-made object involved in an action |
| Auxiliary Party | an actor that in some way participates in an action but is neither the agent nor the target |
| Condition | a constraint stating the properties that must be met |
| Definition | a legal provision defining the meaning of concepts |
| Constraint | a restriction placed on the applicability of a legal provision |
| Exception | a constraint indicating that a legal provision takes precedence over another legal provision |
| Fact | something that is known or proved to be true |
| Location | a place where an action is performed |
| Modality | a verb indicating the modality of the action (e.g may, must, shall) |
| Obligation | a provision imposing mandatory action to be performed by an agent |
| Penalty | a provision indicating the result of breaking an obligation or prohibition |
| Permission | a provision indicating the possibility to perform an action without an obligation or a prohibition |
| Prohibition | a provision forbidding an action to happen or take place |
| Reason | the rationale for an action |
| Reference | a mention of other legal provision(s) or legal text(s) affecting the current provision |
| Result | the outcome of an action |
| Sanction | a punishment imposed in a penalty |
| Statement | a (well-formed) sentence within a legal text |
| Situation | a description of something that has happened or can happen |
| Target | an entity that is directly affected by an action |
| Time | the moment or duration associated with the occurence of an action |
| Violation | a condition that indicates explicit criteria for non-compliance (denial of a provision) |

*Definition* is a legal provision that defines the meaning of a concept. We adopted this statement-level concept from Breaux et al.'s upper ontology (Breaux and Antón, 2008) as we found it in other classifications (Massey et al., 2010; Maxwell and Antón, 2010) as well. This is also aligned with the concept of constitutive statement that is present in the Knowledge Representation community (Grossi et al., 2008).

*Obligation*, *Permission* and *Prohibition* are three modal statement-level concepts that can be found across the legal literature at different granularity levels, as we previously described in Section 2. Traditionally, these statements are related to the use of modal verbs. For example, in *permission* statements such as Statements 1 and 3 in Fig. 1, the "may" modality induces the statement-level concept *permission*. However, in French, modal verbs are not always used for stating obligations or prohibitions, and other cues are needed. For example, the statement "the police officer notifies the driver ..." describes an implicit obligation for the police officer to perform an action.

Finally, we adopted the statement-level concept *penalty* from the Knowledge Representation community. A penalty statement is a provision that imposes sanctions in case of non-compliance. Statement 2 in Fig. 1 provides an example of a penalty statement.

We now continue the presentation of our conceptual model with phrase-level concepts.

*Agent* is an actor performing an action, whereas *target* is an actor affected by the action stated in the provision. A third form of actor is *auxiliary party*, which is neither an agent nor a target, but rather an intermediary. Examples of agents and targets are provided in Statements 1 and 2, respectively. An example of co-occurrence of all three actor types is provided in Statement 3.

The concept of *artifact* captures human-made objects (physical or virtual). An example is "the agreement" in Statement 2.

The concept of *situation* describes a state of affairs, similarly to Nomos (Siena et al., 2012). A situation may be a *result*, and a result may in turn be classified as a *sanction*. An example is "the prohibition of driving" in Statement 3.

The description of "what is happening" is considered as a norm in Nomos (Siena et al., 2012), an action in GaiusT (Zeni et al., 2015), an act in Breaux et al.'s upper ontology (Breaux and Antón, 2008), and a clause in LegalGRL (Ghanavati et al., 2014). In our model, we follow GaiusT's terminology thus adopting the term *action*. As illustrated by our statements in Fig. 1, an *action* can be associated to a *modality* (often expressed via a modal verb) and to *constraints*. Constraints may be further classified as *exceptions* or *conditions*. Conditions may in turn be classified as *violations*, when they describe the circumstances under which the underlying statements are breached (violated). Statement 2 provides an example of violation. Violations, alongside sanctions discussed earlier, provide information that is necessary for inferring the consequences of non-compliance.

We capture the purpose of a statement using the concept of *reason* (not illustrated in Fig. 1). This concept corresponds to *purpose* in Breaux et al.'s upper ontology, to *goal* in GaiusT, and to *reason* in LegalRuleML. Finally, a statement may contain information in the form of *references*, *times* and *locations*. These concepts are all illustrated in the statements of Fig. 1.

As a final remark, we note that not all the concepts discussed in Section 2 have been retained in our model. The decision not to retain was made when we deemed that a concept could be expressed using other concepts, or when the concept could not be directly captured as metadata. For example, *compliance* results from the satisfaction of one or more conditions; *delegation* is a particular type of action involving an auxiliary party; *exclusion* is an implicit type, difficult to infer without additional reasoning.

## 4 Qualitative Analysis of legal Concepts

In this section, we report on the qualitative study aimed at defining extraction rules for semantic legal metadata.

***Study context and data selection.*** We conducted our study in collaboration with Luxembourg's Central Legislative Service (*Service Central de Législation*, hereafter SCL). SCL's main mandate is the publication and dissemination of national legal texts. SCL already employs a range of semantic web technologies for legal text processing, and has considerable prior experience with legal metadata. In recent years, SCL has been investigating the use of legal metadata for two main purposes:

1. Assisting IT engineers in identifying legal provisions that are likely to imply software requirements (Section 1 describes several use cases of semantic legal metadata for requirements analysts).
2. Providing an online service that enables lay individuals and professionals alike to interactively query the law, e.g., ask questions such as: "What would be the consequences of driving too fast on a road with a speed limit of 30 km/h?".

Our work is motivated by the former use case.

Our study focuses on the traffic laws for Luxembourg. They consist of 74 separate legal texts, including state-level legislation, regulations, orders and jurisprudence. Collectively, the texts are 1075 pages long and contain ≈12000 statements. The oldest (and main) text is from 1955 and the most recent from 2016.

The choice of traffic laws was motivated by two factors. First, due to these laws being intuitive and widely known, SCL found them convenient for showing the benefits of legal metadata to decision makers in Luxembourg. Second, the provisions in traffic laws are interesting from an RE perspective, due to their broad implications for the IT systems used by the law enforcement agencies, the courts, and the public infrastructure management departments.

Our study is based on 200 randomly selected statements from the traffic laws. As it is the case with most legal texts, the source texts in our study contain statements with enumerations and lists embedded in them. To treat these statements appropriately, we took the common legal text preprocessing measure of merging the beginning of a statement with its individual list items to form complete, independent sentences (Dell'Orletta et al., 2012).

*Analysis procedure.* Our analysis procedure follows *protocol coding* (Saldaña, 2015), a method for collecting qualitative data according to a pre-established theory, i.e., a set of codes. In our study, the codes are the phrase-level concepts of the model of Fig. 3. The first author, who is a native French speaker and expert in NLP, analyzed the 200 selected statements from the traffic laws, and annotated the phrases of these statements. Throughout the process, difficult or ambiguous situations were discussed between the authors (including a legal expert) and decisions were made based on consensus.

To assess the overall reliability of the coding, the second author – also a native French speaker, with background in NLP and regulatory compliance – independently annotated 10% of the selected statements, prior to any discussion among the authors. Inter-annotator agreement was then computed using Cohen's $\kappa$ (Cohen, 1960). Agreement was reported when both annotators assigned the same metadata type to the same span of text. Other situations counted as disagreements. We obtained $\kappa = 0.824$, indicating "almost perfect agreement" (Landis and Koch, 1977).

*Coding results.* The coding process did not prompt the use of any concepts beyond what was already present in the conceptual model of Fig. 3. In other words, we found the concepts of the model to be adequately expressive.

Table 3 presents overall statistics about the studied statements by indicating the occurrences of each type of statement-level and phrase-level concept. In the majority of cases, we were able to assign a unique annotation to a given phrase. However, we noted that in some cases different interpretations of the same phrase would result in different annotations. The last column of the table provides information about such phrases. For instance, we annotated 73 phrases with the unique concept of *artifact*; in addition, we annotated seven phrases as both *artifact* and *sanction*, five phrases as both *artifact* and *situation*, and so on. We note that phrases are hierarchical and nested: consequently, nested annotations are prevalent, as illustrated by the statements in Fig. 1. What we show in the last column of Table 3 excludes nesting, i.e., it covers only phrases where more than one annotation is attached to exactly the same span. An example phrase is "temporarily or permanently" found in Statement 1 of Fig. 1: here two annotations, namely *constraint* and *time*, have been attached to the same span.

In total, we identified 1339 phrases in the 200 selected statements. Of these phrases, 1299 ($\approx 97\%$) have a single annotation, and the remaining 40 ($\approx 3\%$) have two annotations (no case was observed, where more than two annotations were possible).

With regard to the coverage of statement-level concepts, we observed at least nine occurrences of each concept, except for *facts*, for which we have none. In the Luxembourgish legislation, facts mostly concern generic assertions of little value to RE, such as the details of the publication in the official gazette or the contents of the preamble of the legislative act. However, case law is likely to contain more instances of fact, providing a deeper understanding of the law and its interpretation; the concept is thus important for RE, as described by Breaux et al.'s upper ontology (Breaux and Antón, 2008).

Table 3: Metadata Annotations Resulting from Qualitative Study

| Concept | Unique Classification | Multiple Classifications |
|---|---|---|
| Definition | 9 | |
| Fact | 0 | |
| Obligation | 120 | |
| Penalty | 20 | |
| Permission | 36 | |
| Prohibition | 15 | |
| **Subtotal** | **200** | |
| Action | 187 | |
| Agent | 42 | |
| Artifact | 73 | +7 sanctions, +5 situations, +3 times, +1 violation |
| Auxiliary Party | 34 | |
| Condition | 230 | +18 times, +1 violation |
| Constraint | 5 | +1 time |
| Exception | 22 | |
| Location | 52 | |
| Modality | 68 | |
| Reason | 21 | |
| Reference | 111 | |
| Result | 0 | |
| Sanction | 91 | +7 artifacts |
| Situation | 162 | +5 artifacts, +2 times, +2 violations |
| Target | 73 | |
| Time | 90 | +3 artifacts, +18 conditions, +1 constraint, +2 situations |
| Violation | 38 | +1 artifact, +1 condition, +2 situations |
| **Subtotal** | **1299** | **40** |

With regard to the coverage of the phrase-level concepts, we have at least 20 occurrences for each concept, with two exceptions: *constraint* has only five occurrences, and *result* has none. Despite our study not having identified any occurrences of *result*, the concept is still to be considered as important. Feedback from legal experts indicated in fact that there is a gap between *situation* and *sanction*. Consider for instance the following example statement (from outside our qualitative study): "If the defect is fixed, the car is not subject to a new inspection." Here, "the defect is fixed" is a regular *situation* appearing as part of a *condition*. What follows, i.e, "the car is not subject to a new vehicle inspection" is the consequence of the first situation; however, this consequence is not a sanction. *Result* is thus a general notion for consequences that are not sanctions.

As for constraints that are unclassifiable as any of the specializations of *constraint* in the model of Fig. 3, consider the following statement: "Drivers of transport units [...] must observe, *with respect to the vehicles ahead of them*, a distance of at least 50 meters [...]." The italicized segment in this statement restricts the interpretation of distance. This constraint, however, qualifies neither as a *condition* nor as an *exception*.

We next describe the extraction rules that we derived from our qualitative study. We do not derive extraction rules for *result* and *constraint*, since the study did not yield a sufficient number of occurrences for these two concepts.

## 5 Automating the Extraction of Legal Semantic Metadata

In this section, we present the extraction rules that we have derived based on the outcomes of the qualitative study described in the previous section. We start by presenting the rules for phrase-level metadata in Section 5.1, followed by the rules for statement-level metadata in Section 5.2, noting that the rules for statement-level metadata make use of phrase-level metadata. Finally, we present an extension of the NLP rules for classifying the specializations of actor, namely *agent*, *target* and *auxiliary party*. With regard to these, we observed that distinguishing them is highly context-dependent, and we were not able to derive rules that were simple enough and yet accurate. We therefore devised an alternative strategy for this particular classification using ML. We describe this strategy in Section 5.3.

### 5.1 Phrase-level Metadata Extraction Rules.

Table 4 presents the extraction rules that we derived by analyzing the 1339 manual annotations in our study. The rules were iteratively refined to maximize accuracy over these annotations. Our rules cover 12 out of the 18 phrase-level concepts in the model of Fig. 3. The concepts that are not covered are: *result*, *constraint* (both due to the scarce observations, as noted above), the three specializations of *actor*, and *(cross-)reference*.

With regard to *reference*, we made a conscious choice not to cover it in our extraction rules. Legal cross-references are well covered in the RE literature, with detailed semantic classifications already available (Maxwell et al., 2012; Sannier et al., 2016), and so is the automated extraction of cross-reference metadata (Sannier et al., 2017, 2016).

In Table 4, the element highlighted in blue in each rule is the target of annotation for that rule. All rules use constituency parsing, except the rules for *actor*, which use both constituency and dependency parsing. Aside from the rules for *action* and *actor*, all the rules are expressed entirely in Tregex (Levy and Andrew, 2006), a widely used pattern matching language for (constituency) parse trees. The rule for *action* annotates every verb phrase (VP), excluding from the span of the annotation any embedded segments of type *modality*, *condition*, *exception*, and *reason*. Note that, to work properly, the rule for *action* has to be run after those for the four aforementioned concepts.

We do not provide a thorough description of Tregex which is already well-documented (Levy and Andrew, 2006). Below, we illustrate some of our rules to facilitate understanding and to discuss some important technicalities of the rules in general.

Consider Statement 1 in Fig. 1. A simplified parse tree for an excerpt of this statement is shown in Fig. 4. The *condition* annotation in this statement is extracted by the following Tregex rule: PP << (condition marker). This rule matches any prepositional phrase (PP) that contains a condition marker (in our example, the term "limit"). Initial sets of markers for all the concepts in Table 4, including conditions, were derived from our qualitative study on the 200 annotated statements from traffic laws. With these initial sets in hand, we followed different strategies for different concepts in order to make their respective sets of markers as complete as possible. We present these strategies next. Table 5 illustrates the markers for different concepts. We note that the terms in Table 5 are translations of the original markers

Table 4: NLP-based Rules for Extracting Semantic Legal Metadata

| Concept | Rule(s) |
|---|---|
| Action | • VP with modality, condition, exception and reason annotations removed |
| Actor | • subject dependency and NP < (actor marker)<br>• object dependency and passive voice and PP < P $ (NP < (actor marker))<br>• object dependency and active voice and NP < (actor marker) |
| Artifact | • NP < (artifact marker)<br>• NP !<< (violation marker) \| !<< (time marker) \| !<< (situation marker) \| !<< (sanction marker) \| !<< (reference marker) \| !<< (location marker) \| !<< (actor marker) |
| Condition | • Srel << (condition marker)    • Ssub < (condition marker)<br>• PP << (condition marker)<br>• NP < (VPinf  !<< (exception marker) & !<< (reason marker))<br>• NP < (VPart  !<< (exception marker) & !<< (reason marker)) |
| Exception | • Srel << (exception marker)    • Ssub << (exception marker)<br>• NP < (VPart << (exception marker))    • PP << (exception marker)<br>• NP << (P < (exception marker) $ VPinf ) |
| Location | • NP < (location marker) |
| Modality | • VN < (modality marker) |
| Reason | • Srel << (reason marker)    • Ssub << (reason marker)<br>• PP << (reason marker)    • NP < (VPart << (reason marker))<br>• NP << (P < (reason marker) $ VPinf ) |
| Sanction | • NP < (sanction marker) |
| Situation | • NP < (situation marker) |
| Time | • NP < (time marker)    • PP < (P < (time marker)) $ NP |
| Violation | • NP < (violation marker) |

**NP**: noun phrase, **PP**: prepositional phrase, **Srel**: relative clause, **Ssub**: subordinate clause,
**VN**: nominal verb,  **VP**: verb phrase, **VPinf**: infinitive clause, **Vpart**: VP starting with a gerund in French. We also note that, for simplicity, the table provides a single set of markers per concept. In practice, different rules for extracting the same concept use different subsets of markers. For instance, "who" and "whose" are treated as concept markers in the first *condition* rule in Table 4 (Srel << (condition marker)), but not in the other four rules.

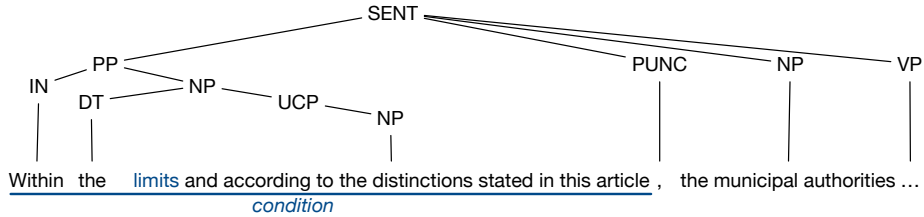

Fig. 4: Simplified Parse Tree for an Excerpt of Statement 1 from Fig. 1

We observed that *actor* and *situation* have broad scopes, thus leading to large sets of potential markers. To identify the markers for these concepts in a way that would generalize beyond our study context, we systematically enumerated the possibilities found in a dictionary. More precisely, we analysed all the entries in Wiktionary (Wiki, 2004). Any entry classified as a noun and with a definition containing "act", "action", or "process" (or variations thereof) counts as a marker for *situation*. For instance, consider the term "inspection", defined by Wiktionary as "*The act of* examining something, often closely". With "inspection" included in the situation markers, the rule for *situation* in Table 4, NP < (situation marker) would mark the noun phrase "vehicle inspections" in Statement 2 of Fig. 1 as a *situation*.

Table 5: Markers for Different Metadata Types

| Concept | Examples of Markers (Non-exhaustive) |
|---|---|
| Fact | |
| Definition | *include, exist, comprise, consist, designate, ...* |
| Penalty | *is comdemned, is punished, is punishable, pronounces, is liable, ...* |
| Permission | *may, can, is permitted, is authorized, ...* |
| Obligation | *shall, must, is obliged, is compelled, is required, ...* |
| Prohibition | *can not, is forbidden, is prohibited, is illegal, is not authorized, ...* |
| Actor* | *physician, expert, company, judge, prosecutor, driver, officer, inspector, ...* |
| Artifact§ | *document, agreement, certificate, licence, permit, warrant, pass, ...* |
| Condition† | *if, in case of, provided that, in the context of, limit, who, whose, which ...* |
| Exception† | *with the exception of , except for, derogation, apart from, other than, ...* |
| Location‡ | *site, place, street, intersection, pedestrian crossing, railway track* |
| Modality† | *may, must, shall, can, need to, is authorized to, is prohibited from, ...* |
| Reason† | *in order to, for the purpose of, so as to, so that, in the interest of, in view of, ...* |
| Sanction† | *punishment, jail sentence, imprisonment, prison term, fine, ...* |
| Situation* | *renewal, inspection, parking, registration, deliberation, ...* |
| Time† | *before, after, temporary, permanent, period, day, year, month, date, ...* |
| Violation† | *offence, crime, misdemeanor, civil wrong, infraction, transgression, ...* |

\* The markers are not generic but are automatically derivable from a simple dictionary.
§ The markers are not generic but can be derived automatically if an ontology like
WordNet's with an explicit classification of objects (human-made and natural) is available.
† The markers are mostly generic and expected to saturate quickly.
‡ The markers are in part domain-specific. Domain-specific markers need to be specified
by subject-matter experts or be derived from an existing domain model (ontology).

In a similar way, any Wiktionary entry classified as a noun and with a definition containing "person", "organization", or "body" (or variations thereof) counts as a marker for *actor*. For example, "authority" is an actor marker since Wiktionary defines it as "A *body* that enforces law and order [...]." As shown by the rules in Table 4, the mere presence of an actor marker does not necessarily induce an *actor* annotation: the candidate phrase must also appear in a subject or object dependency as defined by the rules. To illustrate, let us again consider Statement 1 of Fig. 1. A simplified dependency graph for an excerpt of this statement is provided in Fig. 5. Here, the actor annotation is extracted by the rule: subject dependency and NP < (actor marker). This rule classifies a noun phrase as an *actor* if the noun phrase contains an actor marker and if it has a subject dependency (*nsubj*) to the main (root) verb within the statement.

For *artifacts*, we need the ability to identify human-made objects. It is possible to develop generalizable automation for this purpose in the English language, with support from ontologies such as WordNet (Princeton University, 2010), providing a classification of objects. In place of such an ontology for French, we derived an initial set of markers from the 200 statements in our qualitative study. We then enhanced these markers by inspecting their synonyms in a thesaurus, and retaining what we found relevant. In addition, we implemented a heuristic (the second rule under *artifact* in Table 4) which classifies as *artifact* any noun phrase that is otherwise unclassifiable.

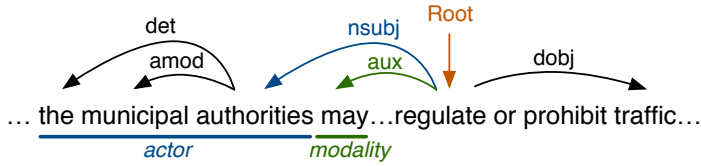

Fig. 5: Simplified Dependency Graph for an Excerpt of Statement 1

Table 6: Rules for Extracting Statement-Level Semantic Legal Metadata

| Statement Concept | Rule | Priority |
|---|---|---|
| Fact | Statement < (Fact Marker) | 6 |
| Definition | Statement < (Definition Marker) | 5 |
| Penalty | Statement < (sanction) OR Statement < (violation) OR Statement < (Penalty Marker) | 1 |
| Permission | Statement < (modality < (Permission Marker)) | 2 |
| Obligation | Statement < (modality < (Obligation Marker)) | 3 |
| Prohibition | Statement < (modality < (Prohibition Marker)) | 4 |

For *conditions*, *exceptions*, *modalities*, *reasons*, *sanctions*, *times*, and *violations*, the markers were derived from our study and later augmented with simple variations suggested by legal experts. As one can see from Table 5, noting the nature of the markers for these seven concepts, the number of possibilities is limited. While our qualitative study did not necessarily capture all the possibilities, we anticipate that the list of markers for these concepts will saturate quickly if enriched further.

Finally, and with regard to the markers for *location*, we followed the same process as described above for *artifacts*, i.e., we derived an initial set of markers from the qualitative study and enhanced the results using a thesaurus. The resulting markers for *location* contain a combination of generic and domain-specific terms. For example, "site" and "place" are likely to generalize to legal texts other than traffic laws. In contrast, designating a "railway track" as a location is specific to traffic laws. The markers for *location* will therefore need to be tailored to the specific legal domains.

5.2 Statement-level Metadata Extraction Rules

Table 6 presents the rules that we have developed for classifying the statements as *facts*, *definitions*, *obligations*, *penalties*, *permissions* or *prohibitions*. The third column indicates the order in which the rules are applied. This ordering is meant at avoiding additional incorrect statement classifications when several rules apply to the same statement.

As we already explained in Section 3, it is quite common to rely on modal verbs like "shall" or "may" to determine the actual deontic nature of a legal statement (Breaux et al., 2006; Zeni et al., 2015; Ghanavati et al., 2014). Therefore, the classification of the three metadata types *obligation*, *permission* and *prohibition* is highly related to the phrase-level metadata *modality*. Since the use of modal verbs is not systematic in French, other cues or markers (see Table 5) are required, notably the interpretation of the main verb of the statement and the implicit modality it implies.

For *penalty*, we mainly rely on the presence of the metadata types *sanction* and *violation*. There also exist markers that directly indicate a *penalty* (see Table 5). For instance, cues such as "punishable" – and its synonyms – are good indicators.

For the remaining two metadata types, namely *fact* and *definition*, we found no obvious markers and therefore the classification relies on the interpretation of the main verb of the statement. From the initial observations, we initiated two lists of markers (one for *fact* and one for *definition*). These lists are limited to the markers found in the qualitative study and their synonyms, and are therefore not exhaustive. We expect the list to saturate quickly over a limited set of qualitative studies on legal acts. For example, a main verb such as "include" can be a good indicator for a *definition*.

In addition to the rules, we also devised a prioritization among the different statement-level concepts. Indeed, *penalty* statements may also have a modal verb such as "may" or "shall", but the presence of a sanction and a penalty cue plays a decisive role in classifying the statement as a *penalty*. Therefore, *penalty* has a higher priority than *obligation*, *permission*, and *prohibition*. An example of such prioritization being triggered for a penalty can be found in Statement 2 of Fig. 1. Finally, if none of the previous rules apply, we attempt to trigger the rules for *definition* and *fact*.

During the qualitative study, we observed that obligations are the most common statements in legal texts. Often (especially in French), the modal verb for expressing obligations is left implicit and therefore cannot be detected by NLP parsers. For this reason we took a design decision to classify by default as *obligation* all the declarative statements that do not contain any cue or marker previously discussed.

5.3 Actor's Role Extraction using Machine Learning

The rules we presented earlier do not account for the roles (subtypes) of *actor*, namely *agent*, *target* or *auxiliary party*. When analyzing a statement, a human annotator is able to further classify actors into its subtypes, thanks to their knowledge of the language and their ability to consider various semantic characteristics of sentences as well as the phrases in the proximity of the actors themselves. Translating such knowledge into intuitive rules, similar to those described above, is possible only for simple and straightforward statements. However, the task becomes challenging when facing the subtleties and complexity of legal language. Nevertheless, an ML algorithm can learn the logic applied by the experts by extracting, from a set of examples, a combination of linguistic elements that a "simple" NLP extraction rule would fail to capture. A good candidate to determine these three *actor* roles is semantic role labeling (SRL) (Gildea and Jurafsky, 2000). SRL aims at automatically finding the semantic role for each argument of each predicate in a sentence. However, the intrinsic limitation of SRL is that it is built to work on relatively simple sentences (Jurafsky and Martin, 2000), and is thus not adapted to long and complex sentences such as those that can be found in legal texts. A second limitation is that the labeling of the inferred roles is based on the analysis of the main verb and provides specific semantic roles (e.g., the agent for the main verb "buy" is labelled "buyer"), whereas our conceptual model revolves around three general concepts (*agent*, *target*, *auxiliary party*). Besides, semantic role labelling frameworks for French such as Verbenet (Pradet et al., 2014) are not likely to yield accurate results for legal texts. Given the arguments above, building an ML-based classifier is a natural course of action. We also want to investigate if we can encapsulate the actor's role extraction into a set of features that an ML technique could combine to yield correct classifications.

*Feature selection.* To train our classifiers, we rely on 31 features, grouped under three categories. These features are shown in Table 7 and described below. They are derived from the linguistic characteristics of the sentence as well as from phrase- and statement-level annotations.

*Sentence-level features.* This set of five features provides information about the statement itself. The first feature is the (active or passive) voice of the main verb of the statement, which has influence over the identification of the semantic subject(s) and object(s) of the statement. The second feature concerns another aspect of the main verb of the statement, namely its transitivity. A verb can be transitive, intransitive or both. This information is extracted from an open source dictionary for French (the French version of Wiktionary) and gives information about the subject and potential object. The third feature returns the modal verb if the statement contains one. A modal verb can be categorized as a marker for permission, obligation or prohibition. The list of modal verbs was manually constructed from the ground truth, and validated by legal experts. An example of such list for obligations is available in Annex A. This feature returns an enumeration of the modal verbs found in the statement, or NULL if the statement does not contain any. The fourth feature is the number of actors in the statement. This information is used in combination with the fifth feature, which identifies the actor that is analyzed in the remainder of the features by indicating its position within the list provided by the fourth feature. When combined with the other features, this indication can provide insights over the role of the actor under analysis.

*Actor's Neighborhood features.* This set of four features categorizes the neighborhood relations of the actor to be classified. The first feature specifies if the actor is contained within another phrase-level annotation. This feature is based on the results of our qualitative study, which showed that agents are usually not contained in other metadata types, while targets are usually contained within the action, and auxiliary parties are easily recognizable if found inside a condition, exception, reason or reference. The second and third features concern the preceding and following phrase-level annotations, respectively. Our qualitative study showed in fact that an empty actor neighbourhood would hint to an agent classification. The fourth feature specifies the type of the preceding POS tag in the sentence (e.g., a preposition). This makes it possible to exclude the classification as agent of actors that are preceded by a proposition.

*Main-verb Relationship features.* This set of 22 features categorizes the relationship between the actor to be classified and the main verb of the legal statement. The first feature is the distance, in terms of number of annotations, between the actor to be classified and the main verb of the statement. For a legal statement, the main verb is recognized through dependency parsing, as previously explained. The next feature is the dependency chain connecting the actor to the main verb, expressed using the information from the dependency graph. This results in an ordered list of dependency types. The remaining 20 features concern the number of instances of a given dependency type in the dependency chain. These dependencies are those that are relevant and that we can extract through the dependency parser. They are similar to the dependencies used, for instance, for Verbenet (Pradet et al., 2014), but adapted to our own parser (the implementation details are elaborated in Section 6).

The presence or absence of a given dependency type and its number of occurrences in the dependency chain, in combination with other features from the actor's neighborhood, can direct the classifier to classify the actor as agent, target, or auxiliary party. For example, the presence of a subject dependency in the dependency chain attached to the *actor* annotation, in addition to an active voice from the sentence level features, is likely to trigger the classification of the actor in question as *agent*. On the other hand, the presence of a multitude of dependency types with a high number of occurrences of modifiers and object dependencies might lead to assigning the role *auxiliary party* to the actor.

Table 7: Classification Features

| Feature Group | Feature | Description |
|---|---|---|
| Sentence information | active voice | indicates the voice of the statement (passive (false) or active(true)) |
| | transitivity of the main verb | indicates the transitivity of the main verb as determined from Wiktionary {'transitive'; 'intransitive'; 'both'} |
| | modal verb | returns the modal verb if retrieved from a modal verb list, 'null' otherwise |
| | number of actors | number of actors in the statement |
| | position of the actor | position of the actor in the list of actors |
| Actor Annotation Neighborhood | Annotation Container | indicatesthe annotation that contains the Actor annotation, 'null' otherwise |
| | Preceding Annotation | indicates the annotation that precedes the actor annotation |
| | Following Annotation | indicates the annotation that follows the actor annotation |
| | Preceeding POS tag | POS tag that preceedes the actor annotation |
| Main Verb Relationship | distance to the main verb | distance in terms of annotations to the main verb |
| | dependency chain | dependency chain from the actor to the main verb |
| | SUJ | number of instances of the dependency subject in the dependency chain |
| | OBJ | number of instances of the dependency direct object in the dependency chain |
| | ATS | number of instances of the dependency predicative complement of a subject in the dependency chain |
| | ATO | number of instances of the dependency predicative complement of a direct object in the dependency chain |
| | MOD | number of instances of the dependency modifier or adjunt in the dependency chain |
| | A-OBJ | number of instances of the dependency indirect complement introduced by à in the dependency chain |
| | DE-OBJ | number of instances of the dependency indirect complement introduced by de in the dependency chain |
| | P-OBJ | number of instances of the dependency indirect complement introduced by another preposition in the dependency chain |
| | DET | number of instances of the dependency determiner in the dependency chain |
| | DEP | number of instances of the dependency in the dependency chain |
| | PONCT | number of instances of the dependency punctuation in the dependency chain |
| | ROOT | number of instances of the dependency root in the dependency chain |
| | DEPCOORD | number of instances of the dependency coordination in the dependency chain |
| | COORD | number of instances of the dependency coordination in the dependency chain |
| | AUXPASS | number of instances of the dependency passive auxiliary in the dependency chain |
| | AUXCAUS | number of instances of the dependency causative auxiliary in the dependency chain |
| | AUXTPS | number of instances of the dependency tense auxiliary in the dependency chain |
| | AFF | number of instances of the dependency affix in the dependency chain |
| | ARG | number of instances of the dependency argument in the dependency chain |
| | MODREL | number of instances of the dependency relative modifier in the dependency chain |

*Dataset.* A key consideration in ML-based classification is the size of the dataset that is needed for training a classifier. In addition to the initial 200 statements of our qualitative study, which contained 149 actors (including 42 agents, 34 auxiliary parties, and 73 targets – see Table 3), we annotated 503 additional statements in order to obtain an enhanced dataset of 1000 actors. This enhanced dataset is composed of 183 agents, 481 targets, and 336 auxiliary parties.

*Training.* We use our dataset to train three classifiers to classify *actors* as *agent*, *target*, and *auxiliary party*, respectively. We use WEKA, which is an open source platform for data mining activities (Frank et al., 2016). We employ ten-fold cross validation over our dataset to evaluate prediction performance of our classifiers. We evaluate several classification algorithms including Naive Bayes classifier, Decision Trees, Random Forest, and Support Vector Machine (SVM). In order to effectively identify the most suitable ML algorithm and optimize the hyper-parameter settings, we run Auto-WEKA (Thornton et al., 2013), a tool to automatically select the optimal classification algorithm among those implemented in the WEKA package using Bayesian optimization.

Overall, Naive Bayes did not yield accurate predictions. This was expected, since Naive Bayes treats features as independent, whereas our features are strongly interconnected. Similarly, Decision Trees did not perform well in our classification task. SVM is not suited for a mix of numerical and categorical features, as it is the case in our dataset. Random Forest (RF) returned actor classifications of a much higher quality than the other ML techniques evaluated by Auto-WEKA. Overall, the ten-fold cross-validation of the RF classifiers led to accuracy results, i.e., the ability to correctly classify annotations as being or not of a given type, of 90.2% for the agent classifier, 79.0% for the target classifier and 78.3% for the auxiliary party classifier, respectively.

*Making the final classification decision.* In our approach, each actor annotation in a statement is submitted for classification over the three classifiers. The final classification decision algorithm is described in Fig. 6. Essentially, our final classification decision favors the highest confidence score, with some additional heuristics being applied:

1. the classifier will favor agent if its confidence score for agent is high;
2. a direct classification is made if the best score is already high and there is a sufficient difference between the best and second best classification scores;
3. when having to choose between *target* and *auxiliary party*, if the difference between the two scores is small, then the classifier will not make a decision ("cannot_classify").

Our heuristics are based on the accuracy measured during the initial cross-validation step, which makes us confident about the classifier for agent, but less so about the classifiers for target and auxiliary party. These two roles are in fact often ambiguous and therefore hard to distinguish. For instance, in statement 3 of Fig. 1, the roles of the public prosecutor and the sued person are not straightforward to establish.

## 6 Tool Support

*Implementation.* Our metadata extraction rules are implemented using Tregex (Levy and Andrew, 2006) and Java. These rules utilize the outputs of the classic NLP pipeline for syn-

**Alg. 1:** Generate an actor classification

**Inputs:**

(1) an actor_annotation to be classified,
(2) SA is the score returned from the agent classifier;
(3) ST is the score returned from the target classifier;
(4) SAux is the score returned from the auxiliary party classifier;
(5) an acceptance threshold T1 for SA;
(6) an acceptance threshold T2 for SA when SA is not the maximum confidence score among the three scores;
(7) an acceptance threshold T3 for uncertainty when we cannot make a classification decision;

**Output:** a classification of the actor annotation

**function GenerateClassification**.
**if** *Agent_Acceptance* **then return** "Agent" .
**else if** *Uncertainty_Condition* **then return** "Cannot_Classify"
**else if** *Target_Acceptance* **then return** "Target".
**else return** "Auxiliary_Party"
**end function**

**function Agent_Acceptance** *// generates a Boolean to decide if the actor annotation should be classified as an agent.*
**return** (SA>T1 or SA==S1 or (SA==S2 and (S1−S2<T2))).
**end function**

**function Target_Acceptance** *//generates a Boolean to decide if the actor annotation should be classified as a target.*
**return** ST == S1.
**end function**

**function Uncertainty_Condition** *// generates a Boolean to decide if the actor annotation can not be classified among the three classes*
**return** S1 − S2 < T3
**end function**

**function S1** *// returns the best confidence score among the three classification models.*

**function S2** *// returns the second-best confidence score among the three classification models.*

Fig. 6: Final Classification Decision Algorithm

tactic analysis. The pipeline described in Fig. 7 has the following modules: Tokenizer, Sentence Splitter, POS Tagger, Named-entity Recognizer, and (Constituency and Dependency) Parser.

Alternative implementations exist for each of these modules. We instantiate the pipeline using a combination of module implementations which we found to be the most accurate for the language of the legal texts in our context, namely French. For the lexical analysis modules, we use standard libraries from Python for the Tokenizer and the Sentence Splitter. With regard to POS Tagging, we use a language-specific framework called Lefff (Sagot, 2010), while we base our Named-Entity Recognition on our specific sets of markers. For constituency and dependency parsing we use the Berkeley Parser (Petrov et al., 2006) and the Malt Parser (Nivre et al., 2007) respectively.

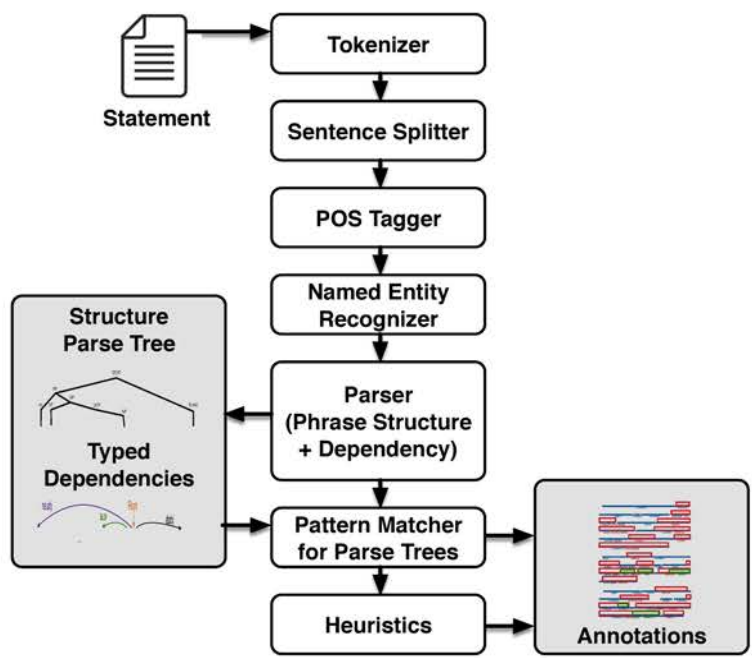

Fig. 7: Tool support

***Design heuristics.*** The heuristics that we have developed (such as wrapping together annotations of the same type that are next to each other, or prioritizing classifications) are related to strategies for avoiding the over-generation of annotations.

The first heuristic is related to the presence of two overlapping annotations of the same type. For example, consider the annotation "the director of the grand-ducal police". In this example, we have two actor markers, one for "the director" and one for "the grand-ducal police". Instead of generating two annotations ("the director of the grand-ducal police" and "the grand-ducal police"), we only consider the one with the longest span, since it identifies more accurately the entity that plays a role in the statement.

The second heuristic is similar to the previous one. It is related to the presence of annotations of different types, where one contains the other. Annotations can contain other annotations, (e.g., a *condition* annotation can include an *actor* annotation). However, annotations contained in *references* do not make sense for the statement analysis. Take for example the "Directive 2014/65/EU of the European Commission...". In this case, "the European Commission" should not be annotated as an *actor*, since it is not related to the action expressed in the statement. We discard annotations of type *actor*, *time*, *location*, *artifact* and *condition* when they are contained within a *reference* annotation.

The third heuristic is related to hierarchically-ordered annotations spanning the same content. For instance, the annotation "an imprisonment" is annotated as both a *situation* and a *sanction*. However, following our previously described conceptual model, *sanction* is a kind of *situation*. In these cases, we discard the annotation of the most generic type (in this example, *situation*).

The last heuristic is related to ambiguity. For instance, let us consider the marker "court of justice": it can mean either the authority (*actor*) delivering the judicial decision, or the *location* where such decision is taken. While a human is able to infer the correct meaning from the general context, automating this task is very difficult since in many cases the sentence does not provide any linguistic cue to help with the disambiguation. Consequently, we decided to assign the ambiguous marker to both metadata types. Although this decreases the overall precision, this marker at least restricts the choice to two metadata types instead of leaving all the 18 classification possibilities open for that particular phrase.

## 7 Empirical Evaluation

In this section, we measure the accuracy of our extraction rules through two case studies.

### 7.1 Research Questions

Our evaluation is targeted at answering the following research questions (RQ):

***RQ1. How accurately can our approach extract semantic legal metadata when it is employed in the same domain as our qualitative study?*** This RQ aims at evaluating the completeness of our extraction rules, markers, and classification techniques, for the extraction of both statement-level and phrase-level metadata when applied to statements in our initial analysis domain, i.e., the traffic laws.

***RQ2. How accurately can our approach extract semantic legal metadata when it is employed outside the domain where we conducted our qualitative study?*** This RQ is an attempt toward evaluating the generalizability of our approach by applying it to different domains within the Luxembourgish legislation.

### 7.2 Case Studies Description

To answer our research questions, we have set up two different case studies that we describe below.

*Case study 1 (CS1).* The objective of this case study is to measure the accuracy of the extraction rules of Table 4 against a ground truth. To build the ground truth, we manually annotated 150 randomly selected legal statements from the traffic laws, in addition to the 200 statements previously annotated for our qualitative study of Section 5. We followed the same protocol coding process as described in our qualitative study. The construction of the ground truth took place strictly after the conclusion of our qualitative study. Specifically, our extraction rules (including the concept markers) were already finalized and frozen at the time when we selected and analyzed the 150 statements. The ground truth was constructed in two rounds. In the first round, we annotated 100 statements and performed a complete round of evaluation, following the same procedure that we explain below. Our analysis of the results in the first round did not lead to new extraction rules, but prompted marginal improvements to the concept markers for *condition*, *time*, and *location* (see Table 5). Following the first evaluation round, we annotated another 50 statements and measured the accuracy of our improved solution over them. We obtained accuracy levels similar to those of the first round. This provides confidence that our extraction rules and markers have saturated. We report the evaluation results for the 100+50=150 statements combined. To avoid biased conclusions, the reported results use the baseline set of concept markers, i.e., the same set with which the first evaluation round was performed.

The first author annotated the 150 statements used in the evaluation, and the second author independently annotated 10% of these statements to examine reliability. We obtained $\kappa = 0.815$, suggesting "almost perfect agreement" (Landis and Koch, 1977). In total, the ground truth has 1202 annotations covering 1177 phrases (25 phrases have double annotations). A detailed breakdown is provided in the ground truth column of Table 8. Similar to the qualitative study, we observed no occurrences of *result* and a very low number of occurrences of *constraint*.

To evaluate our extraction rules, we exclude occurrences of *constraint* and *result* for which we do not provide rules, and occurrences of *reference*, whose detection we leave to existing solutions. Our evaluation is thus based on 1127 ground-truth annotations.

*Case study 2 (CS2).* The objective of this case study is to explore the generalizability of our approach for semantic metatadata extraction in different domains of the legislation, and to investigate how the rules and markers that we have acquired in the qualitative study and in the first case study can expand to other legislative domains. To do so, we randomly selected 40 statements from five Luxembourgish legislative codes, including the Code of Commerce, the Penal Code, the Code for healthcare, the Labor Code and the Code for the Environment, for a total of 200 statements. Among these codes, each of the first two acts is a single harmonized legal act, whereas each of the last three is a collection of legislative acts of various nature (laws, regulations, etc.). These codes possess their own terminology, thus providing an interesting context in which to investigate how important domain knowledge is in the automatic extraction of semantic metadata.

Regarding the manual construction of the second ground truth, the 200 statements were manually annotated by the second author. This is to mitigate the potential bias due to the fact that the implementation of the tool was performed by the first author. Similarly to the first case study, the first author independently annotated 10% of the statement to assess reliability. We obtained $\kappa = 0.813$, suggesting "almost perfect agreement" (Landis and Koch, 1977). In total, the ground truth for this case study has 2132 annotations, covering 1974 phrases (158 phrases have double annotations). The detailed breakdown is provided in the ground truth column of Table 9.

Similarly to the first case study, and for the same reason, we do not report on *constraint*, *result* and *reference*.

7.3 Analysis Procedure

Each phrase-level annotation has two parameters: a *type* and a *span*. The former specifies the legal concept, while the latter specifies where the annotation begins and where it ends in the statement. We evaluate the results of automated phrase-level metadata extraction using the following notions:

– A computed annotation is a *match* if its span has a non-empty intersection with some ground-truth annotation of the same type.
– A computed annotation is *misclassified* if it is not a match.
– A ground-truth annotation for which there is no match is considered as *missed*.

For sentence-level annotations, only the *type* parameter is pertinent; the *span* is implied since the annotation covers the entire statement. We therefore evaluate the results of automated sentence-level metadata extraction using the following notions:

– A computed annotation is a *match* if its type matches that of the ground-truth annotation for the same statement.
– A computed annotation is *misclassified* if it is not a match.
– A ground-truth annotation for which there is no match is considered as *missed*.

Our evaluation results are presented in columns 3 through 8 of Table 8. For each legal concept (metadata type), we provide the number of matches, the number of misclassified annotations, the number of missed annotations, and scores for precision and recall. Each match counts as a true prositive (TP); each misclassified annotation counts as a false positive (FP), and each missed annotation counts as a false negative (FN). Precision is computed as $|TP|/(|TP| + |FP|)$ and recall as $|TP|/(|TP| + |FN|)$.

7.4 Results for the First Case Study

We first discuss the results for the classification of phrase-level metadata, except actors. This is followed by the results for the classification of actors (into sub-types) and statements. The classification of both actors and statements relies on other phrase-level metadata (see Sections 5.2 and 5.3). Finally, we perform an error analysis of the misclassifications and the missed metadata types.

*Results for phrase-level metadata.* In summary, out of the 1100 computed annotations, 1069 annotations were correct matches and 31 (2.8%) were misclassified. There are 58 ground-truth annotations (5.1%) that were missed by the extraction rules, due to either misclassification or the impossibility to classify (i.e., the span of the ground-truth annotation lacks a computed annotation altogether). The impossibility to classify was mostly due to parser errors, or to missing Tregex patterns (or markers) that are either too rare or inconclusive, i.e., they are not unambiguous enough to be used as patterns (or markers) for classification. We obtain an overall precision of 97.2% and an overall recall of 94.9%.This means that our approach identifies the types of metadata items with very high accuracy. Analysts can thus expect to have a correct type assigned automatically in the large majority of cases.

*Results for actor classification.* The results from the automated classification of actors are aligned with what we observed during the training of the classifiers and the 10-fold cross validation. The results are far better than in our initial evaluations based on rules, with increases of $\approx 30\%$ for precision and $\approx 40\%$ for recall when detecting the sub-types of actor. Still, our error analysis highlights that the large majority of issues are related to ambiguities between *targets* and *auxiliary parties*. Consider for example the following complex statement related to driving a car rented in a foreign country (simplified): "Any vehicle belonging to *a physical or moral person* having their main residence or office located in another European State [...] and allowed to perform car leasing can, on the basis of the registration document drawn up by *the competent authorities of that State*, be put into circulation on the public roads of Luxembourg by *any person* having its residence or office in Luxembourg if this vehicle was made available to *this person* through a lease [...]." In this statement we have italicized the different *actor* annotations, namely: "a physical or moral person", "the competent authorities of that State", "any person", and "this person". Here the last annotation ("this person") refers to the previous one ("any person"), and both should therefore be classified as *target*. However, our classifier wrongly considers "any person" to refer to the first two annotations, which are *auxiliary parties*, thus generating a wrong classification for the last two actors.

Table 8: Statistics for Automated Semantic Metadata Extraction (CS1)

| Legal Concept | Ground Truth | Results of Automatic Metadata Extraction | | | Accuracy | |
| --- | --- | --- | --- | --- | --- | --- |
| | | Extracted | Match (TP) | Misclassified (FP) | Missed (FN) | Precision | Recall |
| Definition | 9 | 9 | 9 | 0 | 0 | 100,0% | 100,0% |
| Fact | 0 | -- | -- | -- | -- | N/A | N/A |
| Obligation | 94 | 96 | 94 | 2 | 0 | 97,9% | 100,0% |
| Penalty | 4 | 4 | 4 | 0 | 0 | 100,0% | 100,0% |
| Permission | 26 | 26 | 26 | 0 | 0 | 100,0% | 100,0% |
| Prohibition | 17 | 15 | 15 | 0 | 2 | 100,0% | 88,2% |
| **Subtotal** | **150** | **150** | **148** | **2** | **2** | **98,7%** | **98,7%** |
| Action | 157 | 157 | 155 | 2 | 2 | 98,7% | 98,7% |
| Actor | 138 | 133 | 129 | 4 | 9 | 97,0% | 93,5% |
| *Agent* | *21* | *22* | *19* | *3* | *2* | *86,4%* | *90,5%* |
| *Aux. Party* | *73* | *69* | *59* | *10* | *14* | *85,5%* | *80,8%* |
| *Target* | *44* | *42* | *36* | *6* | *8* | *85,7%* | *81,8%* |
| Artifact | 252 | 241 | 238 | 3 | 14 | 98,8% | 94,4% |
| Condition | 172 | 179 | 166 | 13 | 6 | 92,7% | 96,5% |
| Constraint | 3 | -- | -- | -- | -- | N/A | N/A |
| Exception | 12 | 12 | 10 | 2 | 2 | 83,3% | 83,3% |
| Location | 35 | 34 | 34 | 0 | 1 | 100,0% | 97,1% |
| Modality | 80 | 81 | 78 | 3 | 2 | 96,3% | 97,5% |
| Reason | 23 | 22 | 22 | 0 | 1 | 100,0% | 95,7% |
| Reference | 72 | -- | -- | -- | -- | N/A | N/A |
| Result | 0 | -- | -- | -- | -- | N/A | N/A |
| Sanction | 29 | 25 | 25 | 0 | 4 | 100,0% | 86,2% |
| Situation | 150 | 140 | 137 | 3 | 13 | 97,9% | 91,3% |
| Time | 67 | 63 | 63 | 0 | 4 | 100,0% | 94,0% |
| Violation | 12 | 13 | 12 | 1 | 0 | 92,3% | 100,0% |
| **Subtotal** | **1202*** | **1100** | **1069** | **31** | **58** | **97,2%** | **94,9%** |

*We exclude from our evaluation *constraints* and *references* as noted in the text: we do not have extraction rules for *constraints*; detecting *(cross-)references* is outside the scope of this paper. This leaves us with 1202-75 = 1127 relevant annotations.

Disambiguating *targets* from *auxiliary parties* in those situations would require a representation of domain semantics, e.g., through a taxonomy, which is left for future work and thus not addressed here.

***Results for statement-level metadata.*** In summary, out of the 150 computed statement-level annotations, only two are incorrect, implying two annotations being misclassified (FP) and an equal number of missing (FN) classifications. In both cases, the tool triggered an obligation whereas a prohibition was the choice of the annotator. This is due to the lack of context: to capture those annotations it is necessary to take into account the preceding statements, which were implicitly taken into account by the annotator, but that the tool is unable to handle. For instance, one of the statements began with "It is the same for ...". In the absence of any other cue, the rule triggered the default choice, i.e., *obligation*. The preceding statement was instead a prohibition, but the tool lacked this information.

***Discussion.*** To answer RQ1, the results presented above show that our approach is highly accurate when employed in the same domain as our qualitative study. To determine the root causes for the automation inaccuracies observed, we analyzed the misclassified and missed annotations. Of the 31 misclassifications in Table 8, 20 are related to polysemous concept markers. For example, the term "seizure" is a marker for *sanction*, since the term may refer to the confiscation of a possession. This term may also refer to a medical symptom, in which case it suggests a *situation*. Both senses of the term are used in the traffic laws. When the term is used in the latter sense, our rules generate a misclassified annotation. Three misclassifications arise from complex legalese and are thus unavoidable. The remaining eight misclassifications are due to constituency parsing errors, discussed later.

Of the 58 missed annotations, 25 are related to double annotations in the ground truth. In all these cases, our rules identify one of the two ground-truth annotations, but we still count one FN for each case, since, compared to a human annotator, the rules lack the ability to take multiple perspectives into account. Among the remaining 33 missed annotations, 26 are due to misclassifications, discussed earlier. Five missed annotations result from a pair of distinct ground-truth annotations that our rules grouped into a single annotation, that intersects with both ground-truth annotations. Each of these five cases leads to one match and one missing annotation. The last two missed annotations are caused by constituency parsing errors, discussed later.

## 7.5 Results for the Second Case Study

Similarly to the first case study, our evaluation results for the second case study are presented in columns 3 through 8 of Table 9 and follow the same presentation order. For the second case study, the results of the evaluation are analyzed in comparison with the results of the first case study. By doing so, we attempt to assess how our findings within the legal domain of our qualitative study – which is the same as that of our first case study – would generalize to other legal domains. The error analysis for each part is directly discussed, while a broader discussion is performed at the end of the section.

***Results for phrase-level metadata.*** When tested against the new multi-domain ground truth, the set of general markers and rules that we have developed maintained good results in terms of recall but not in terms of precision. In summary, over the 2213 computed annotations, 1823 annotations were correct matches and 390 (17.6%) were misclassified. Our rules missed 150 ground-truth annotations (7.6%), due to either misclassification or impossibility to classify (due to parser errors, or to missing Tregex patterns or markers). The overall precision and recall for phrase-level annotations reach 82.4% and 92.4%, respectively. The overall recall for this case study (92.4%) is comparable to the first case study (94.9%); nevertheless, we observe a decrease in precision from 97.2% to 82.4%. As the following error analysis and discussion show, this decrease can be attributed to the increased length of the statements (see Table 10) and to the domains being different.

With regard to *artifact*, the lower precision score is related to the absence of markers for other elements, leading to *artifact* as a default classification. In fact, since the domains are different from that of our qualitative study, we expected that our list of general markers would not be able to cover all the new terms.

With regard to *condition*, most of the misclassifications are related to our inability to extract *constraints*, which led to considering them as *conditions* instead. These new occur-

Table 9: Statistics for Automated Semantic Metadata Extraction (CS2)

| Legal Concept | Ground Truth | Results of Automatic Metadata Extraction | | | | Accuracy | |
|---|---|---|---|---|---|---|---|
| | | Extracted | Match (TP) | Misclassified (FP) | Missed (FN) | Precision | Recall |
| Definition | 13 | 11 | 9 | 2 | 4 | 81,8% | 69,2% |
| Fact | 2 | -- | -- | -- | 2 | N/A | N/A |
| Obligation | 114 | 122 | 107 | 15 | 7 | 87,7% | 93,9% |
| Penalty | 30 | 29 | 27 | 2 | 3 | 93,1% | 90,0% |
| Permission | 33 | 28 | 28 | 0 | 5 | 100,0% | 84,8% |
| Prohibition | 8 | 10 | 7 | 3 | 1 | 70,0% | 87,5% |
| **Subtotal** | 200 | 200 | 178 | 22 | 22 | 89,0% | 89,0% |
| Action | 165 | 169 | 151 | 18 | 14 | 89,3% | 91,5% |
| Actor | 526 | 521 | 497 | 24 | 29 | 95,4% | 94,5% |
| *Agent* | 114 | 111 | 99 | 12 | 15 | 89,2% | 86,8% |
| *Aux. Party* | 239 | 251 | 210 | 41 | 29 | 83,7% | 87,9% |
| *Target* | 173 | 159 | 135 | 24 | 38 | 84,9% | 78,0% |
| Artifact | 319 | 461 | 293 | 168 | 26 | 63,6% | 91,8% |
| Condition | 301 | 321 | 252 | 69 | 49 | 78,5% | 83,7% |
| Constraint | 55 | -- | -- | -- | -- | N/A | N/A |
| Exception | 14 | 12 | 11 | 1 | 3 | 91,7% | 78,6% |
| Location | 87 | 99 | 82 | 17 | 5 | 82,8% | 94,3% |
| Modality | 72 | 102 | 71 | 31 | 1 | 69,6% | 98,6% |
| Reason | 18 | 27 | 14 | 13 | 4 | 51,9% | 77,8% |
| Reference | 97 | -- | -- | -- | -- | N/A | N/A |
| Result | 7 | -- | -- | -- | -- | N/A | N/A |
| Sanction | 57 | 58 | 56 | 2 | 1 | 96,6% | 98,2% |
| Situation | 249 | 272 | 246 | 26 | 3 | 90,4% | 98,8% |
| Time | 132 | 145 | 128 | 17 | 4 | 88,3% | 97,0% |
| Violation | 33 | 26 | 22 | 4 | 11 | 84,6% | 66,7% |
| **subtotal** | 2132* | 2213 | 1823 | 390 | 150 | 82,4% | 92,4% |

*see foot note in Table 7. Here, this leaves us with 2132-55-97-7 = 1973 relevant annotations.

rences of *constraint* can however be the basis for the elaboration of appropriate extraction rules for this concept. The missed annotations of *condition* are related to cues that cannot be considered as specific markers as-is, as they need to be associated with a deeper NL analysis of the statement. For instance, phrases such as "by (him)", "within", "(member) of", "under (the name)" can hardly be considered as viable cues for *condition* on their own, since they can also be used in other contexts.

With regard to *exception*, the number of occurrences is low. From the initial observations, it emerged that their classification mostly depends on the context of the preceding statements. Because our approach treats each statement in isolation, it is unable to correctly identify such annotations.

With regard to *location*, issues are related to the ambiguity between the location and the authority (actor) who has its premises in that location, since both are referenced using the same term (e.g., a court of justice, school, or third party country).

The issues for *modality* are related to the improper handling of modal verbs which are not the main verb of the statement. These modal verbs in fact should not be classified as *modality* since they have no effect on the classification of the statement that contains them. For instance, in the statement "the officer notifies the person that she *can* ...", the permission modality "can" does not affect the main verb ("notify"), which instead bears an implicit obligation. Our markers and rules however do not take this into account, leading to mistakes at the statement level.

Concerning *reason*, we only have few occurrences, and a deeper analysis of the structure and the content of the statement is thus required in the future.

With regard to *violation*, we had few observations in the traffic laws, and could therefore only extract a limited set of markers for this metadata type. In addition, many violations are vague and implicit, and are understood as such only because of the context of the statement. For instance, "entering a (restricted) area" (implicitly, without authorization) does not provide sufficient cues about a possible *violation*.

***Results for actor classification.*** The results for the automated classification of actors in the second case study are consistent with the results in the first case study, although we do observe a slight decrease (2%) in recall. Regarding the classification of actors as *agents*, *targets*, and *auxiliary parties*, we note that the evaluation takes into account the actors that we failed to classify as such in the first extraction phase. They are therefore counted as false negatives. Regarding the classification of actors, results for *target* and *auxiliary party* have improved though those for *agent* have worsened.

***Results for statement-level metadata.*** We do not tackle the *fact* type in our statement-level metadata; as noted in Section 5.2, we do not have rules for classifying statements of this type. Our main observation in relation to the remaining statement-level metadata types is that our rules did not maintain precision and recall as high as in the first case study, with both scores decreasing from 98,7% to 89%. When analyzing false positives and false negatives, the following explanations emerge. With regard to *definitions*, we only had few cues from the initial qualitative study and the first case study. We encountered more *definitions* in the second case study, which gave us access to more markers, and this circumstance may improve the results in future iterations. Nevertheless, it often happens that a sentence does not contain unambiguous cues that can lead to its classification as a definition, whereas a human can still interpret it as such. An example is the verb "to be", which is used to express classification (and therefore definition) in sentences such as "Traders are those who buy and sell goods, currency, or shares". In these cases, the classifier fails to identify the statement as a *definition*, but in the presence of more unambiguous verbs such as "consider", which is a viable *definition* marker ("Traders are *considered as* those who buy and sell goods, currency, or shares"), it would have succeeded.

Similarly to the first case study, the classification of *prohibition* was affected by implicit information from a previous statement, which our rules do not account for. With regard to *permission*, the results are affected by a missing marker ("is likely to"), which denotes the possibility of a future action to be performed, as well as by the issue for *modality* when in presence of modal verbs that are not the main verb of the statement (as noted above). Obviously, *penalty* statements are affected by the ability to detect *violation* and *sanction* annotations. Finally, the precision for *obligation* has decreased, which is a direct consequence of the previous misclassifications, since obligation is a default classification in our approach (see Section 5.2).

***Discussion.*** Compared to the first case study, the main reason for the drop in precision is related to the length of the statements. Table 10 shows the average length in words of the statements for our case studies and the standard deviations. We observe a significant length increase in the second case study, where statements tend to be longer, more nuanced and more complex than in the first case study. Here, longer statements are related to the introduction of long cross-references but also due to the presence of conditions and constraints that apply to specific elements in the statements. These additions also tend to break the flow of the statement.

Table 10: Average Statement Length in CS1 and CS2

|                    | Average Statements Length (#Words) | Standard deviation (#Words) |
|--------------------|-----------------------------------:|----------------------------:|
| Traffic Code       | 35.3                               | 21.2                        |
| Code of Commerce   | 56.4                               | 32.3                        |
| Environmental Code | 66.0                               | 42.3                        |
| Penal Code         | 69.9                               | 35.4                        |
| Health Code        | 49.0                               | 34.3                        |
| Labour Code        | 58.0                               | 38.5                        |

As indicated previously, constituency and dependency parsers are more prone to errors when the length of the sentence increases. Beyond $\approx$ 30-35 words, parsers' accuracy starts to decrease rapidly (McDonald and Nivre, 2007; Kummerfeld et al., 2012). In the first case study, the average length of statements is already reaching the upper limit. In the second case study, the average length is significantly above this limit and therefore parsers' accuracy is likely to drop. This is due to the nature of the training data over which existing parsers have been trained. The training data is traditionally extracted from newspaper articles, with relatively short or medium-sized sentences written in a style that is not as formal as in legal texts. Accurate parsing of complex and lengthy legal statements would require dependency and constituency parsers to be trained specifically over legal texts. Labeling a sufficiently large corpus of legal texts for training parsers would necessitate a substantial amount of manual effort and is beyond the scope of this article.

Problems in detecting the markers in the right parse trees may negatively influence the quality of the annotations in the vicinity of the missed marker. This is due to the combination of the second extraction rule for artifact in Table 4 and the first two heuristics presented in Section 6. The extraction rule automatically classifies a phrase-level concept as artifact by default, i.e., if no other classification rule can be applied. The two heuristics prevent the addition of annotations within a span that is already annotated as either actor or artifact, in order to limit the over-generation of annotations. As a consequence, if a marker is missed in the parse tree, the concept will be misclassified as an artifact and any other annotation that is contained within (or overlaps with) the misclassified annotation will be discarded altogether.

Furthermore, nuanced and complex statements are likely to present partial semantics and implicit information, especially when they are the continuation of previous statements from which they borrow part of the context. An example is the statement "The request must be sent to the competent tribunal". In this sentence, we lack the identity of the agent, the type of the request, and the conditions under which such a prescription applies. All this information is located in the same legal text, but in a previous statement. While a human analyst is able to locate such information, our approach is unable to do so because it treats each statement in isolation. Addressing this issue would enable a better detection of metadata when facing incomplete information, and will be part of our future work.

Regarding the influence of domain knowledge and the accuracy of our rules, since we did not perform any qualitative analysis over the new domains in the second case study, our rules and markers are based only on our initial observations from the traffic laws, and on the general knowledge that we have extracted from Wiktionary and other synonyms dictionaries (see Section 5.1). To answer RQ2, the results of the second case study did not prompt the need for modifying or adding new rules (except for *constraint*), and only prompted a limited need for updating the markers for some concepts.

As a final remark, in the absence of explicit markers, classifiers are unable to automatically process legal concepts and legal statements if relying only on the linguistic characteristics of sentences. This is particularly the case for ambiguous or polysemous words, and for statements such as *definitions* and *violations*, when cues may be too generic for use as reliable markers (e.g., "to be" for *definition*) or when a simple situation (e.g., entering a location) has to be interpreted as a violation because of implicit contextual information (e.g., entering a location *without authorization*). For ambiguous or polysemous terms, a specialized domain glossary would be needed. For general or incomplete cues, classification would require a deeper representation of the relationships between the phrase-level metadata, in order to fine-tune the extraction rules for these metadata. This is left for future work.

## 8 Threats to Validity

The most pertinent threats to the internal and external validity of our work are discussed below.

*Internal validity.* A potential threat to internal validity is related to the interpretation of the existing legal metadata types performed by the authors. In order to mitigate the threat posed by subjective interpretation, we tabulated all the concepts identified in the literature and established a mapping between them. By doing so, we helped ensure that no concepts were overlooked, and that the correspondences identified between the metadata types were rooted in the existing definitions. While we cannot rule out subjectivity, we provide our interpretation in a precise and explicit form in Section 3. This is thus open to scrutiny.

Another potential internal validity threat is that the coding in both the qualitative study of Section 5 and the case studies of Section 7 was done by the authors. Since traffic laws are intuitive and one of the authors (i.e., the last author) is a legal expert, we found the risk of misinterpretation during coding to be low. Furthermore, to prevent bias in the coding process, we took several mitigating actions: (1) we carefully discussed the difficult cases encountered during coding; (2) we completed the coding component of our qualitative study before defining any extraction rules; (3) we did not apply our implementation to the legal statements in the ground truth until the coding was completed, in order to minimize the influence of the extraction rules on the construction of the ground truth in the first case study; (4) we assessed the reliability of the coding results by measuring interrater agreement over 10% of the coded statements. With regard to the second case study, the ground truth was built by the second author in order to limit the influence of the implementation. Again, we evaluated the coding work by having another annotator (the first author) code a sample (10%) of the statements, and measuring inter-annotator agreement.

*External validity.* The nuanced nature of legal texts often requires research on legal requirements to be based upon qualitative results obtained in specific contexts. However, a qualitative study with a scope as limited as ours makes it difficult to address external validity with sufficient rigor. Although we have extended our case studies to new domains with promising results, this extension also showed a difference in the results, which is related to the complexity of the statements. Therefore, further studies that still cover a variety of legal domains and in larger settings remain essential for ascertaining the general applicability of our results.

With this said, the following observations provide a degree of support for the external validity of our qualitative study: First, the rules of Table 4 are, in general, simple; there is no particular reason to suspect that these rules may be domain-specific. This helps mitigate the risk of overfitting the rules to our study context. The second case study did not prompt the need for modifying the rules that we already had, and enabled us to add rules and markers for *constraint*, for which we did not have any. This improves our confidence in the meaningfulness of our results across different domains. Second, as we argued while discussing the concept markers of Table 5, most of the marker sets are either systematically extractable from existing lexicons, or expected to saturate quickly due to the limited set of possible linguistic variations. As shown in the second case study, our markers generalized well to other legal domains. However, the study also prompted the need for a more nuanced set of markers, able to handle polysemous and vague terms that can pose a challenge in the interpretation of a statement. We note that the markers are necessarily language-dependent and do not carry over from one language to another.

## 9 Conclusion

Automatic extraction of metadata about the semantics of legal statements is an important enabler for legal requirements analysis. In this article, we first described an attempt at reconciling the different types of semantic legal metadata proposed in the RE literature. We then derived, through a qualitative study of traffic laws, extraction rules for the reconciled metadata types. Our extraction rules are based on natural language processing, more precisely on constituency and dependency parsing, and are complemented with machine learning for distinguishing subtypes of the metadata type *actor*. We evaluated our extraction rules via two case studies, covering 150 statements from the traffic laws and 200 statements from five different legislative domains (commerce, environment, health, labour, and penal codes). The results are promising: we obtained a precision of 97.2% and 82.4% and a recall of 94.9% and 92.4% for the first and second case studies respectively. The loss of precision in the second case study is mainly related to the increased length of the statements, which challenges the ability of existing NLP parsers to perform accurately. Still, our results give us confidence in the ability of our rules and markers to achieve good accuracy across different domains.

In the future, we plan to perform a deeper analysis of our produced metadata. Currently, apart from some prioritization in the classification, each metadata is extracted independently of other adjacent metadata. It would be interesting to see if the analysis of the relationships between metadata types can help achieve a more accurate interpretation of the phrases, resulting in better metadata. Another direction is to analyze the usefulness and relevance of these semantic legal metadata in a practical use case, such as a query system to answer the questions that a requirements analyst could ask when elaborating legal requirements. We have already started investigating this direction in (Sleimi et al., 2019), but further work remains, e.g., semantic integration with cross-reference resolution (Sannier et al., 2017), and identifying the true subject of a given legal statement.

***Acknowledgments.*** Supported by the Luxembourg National Research Fund (FNR) under grants PUBLIC2-17/IS/11801776 and PoC16/11554296.

**References**

Arora C, Sabetzadeh M, Briand LC, Zimmer F (2015) Automated checking of conformance to requirements templates using natural language processing. IEEE Transactions on Software Engineering 41(10):944–968

Athan T, Boley H, Governatori G, Palmirani M, Paschke A, Wyner AZ (2013) OASIS LegalRuleML. In: Proceedings of the International Conference on Artificial Intelligence and Law (ICAIL'13), pp 3–12

Bhatia J, Breaux TD, Schaub F (2016a) Mining privacy goals from privacy policies using hybridized task recomposition. ACM Transactions on Software Engineering and Methodology 25(3):22:1–22:24

Bhatia J, Evans MC, Wadkar S, Breaux TD (2016b) Automated extraction of regulated information types using hyponymy relations. In: Proceedings of the 3rd International Workshop on Artificial Intelligence for Requirements Engineering (AIRE'16), pp 19–25

Boella G, Caro LD, Humphreys L, Robaldo L, Rossi P, van der Torre L (2016) Eunomos, a legal document and knowledge management system for the web to provide relevant, reliable and up-to-date information on the law. Artificial Intelligence and Law 24(3):245–283

Boer A, Winkels R, Vitali F (2007) Proposed XML standards for law: Metalex and LKIF. In: Proceedings of the 20th Annual Conference on Legal Knowledge and Information Systems (JURIX'07), pp 19–28

Breaux T (2009) Legal requirements acquisition for the specification of legally compliant information systems. PhD thesis, North Carolina State University, Raleigh, North Carolina, USA

Breaux TD, Antón AI (2008) Analyzing regulatory rules for privacy and security requirements. IEEE Transactions on Software Engineering 34(1):5–20

Breaux TD, Vail MW, Antón AI (2006) Towards regulatory compliance: Extracting rights and obligations to align requirements with regulations. In: Proceedings of the 14th IEEE International Requirements Engineering Conference (RE'06), pp 46–55

Breuker J, Boer A, Hoekstra R, van den Berg K (2006) Developing content for LKIF: ontologies and frameworks for legal reasoning. In: Proceedings of the 19th Annual Conference on Legal Knowledge and Information Systems (JURIX'06), pp 169–174

Cohen J (1960) A coefficient of agreement for nominal scales. Educational and Psychological Measurement 20(1)

Dell'Orletta F, Marchi S, Montemagni S, Plank B, Venturi G (2012) The splet2012 shared task on dependency parsing of legal texts. In: the 4th Workshop on Semantic Processing of Legal Texts (SPLeT'12), pp 42–51

Elrakaiby Y, Ferrari A, Spoletini P, Gnesi S, Nuseibeh B (2017) Using argumentation to explain ambiguity in requirements elicitation interviews. In: Proceedings of the 25th IEEE International Requirements Engineering Conference (RE'17), pp 51–60

Evans MC, Bhatia J, Wadkar S, Breaux TD (2017) An evaluation of constituency-based hyponymy extraction from privacy policies. In: Proceedings of the 25th IEEE International Requirements Engineering Conference (RE'17), pp 312–321

Frank E, Hall MA, , Witten IH (2016) The WEKA workbench. online appendix for "data mining: Practical machine learning tools and techniques"

Ghanavati S (2013) Legal-urn framework for legal compliance of business processes. PhD thesis, University of Ottawa, Ottawa, Ontario, Canada

Ghanavati S, Amyot D, Rifaut A (2014) Legal goal-oriented requirement language (legal GRL) for modeling regulations. In: Proceedings of the 6th International Workshop on

Modeling in Software Engineering (MISE'14), pp 1–6

Gildea D, Jurafsky D (2000) Automatic labeling of semantic roles. In: the 38th Annual Conference of the Association for Computational Linguistics (ACL-00), pp 512–520

Giorgini P, Massacci F, Mylopoulos J, Zannone N (2005) Modeling security requirements through ownership, permission and delegation. In: Proceedings of the 13th IEEE International Conference on Requirements Engineering (RE'05), pp 167–176

Gordon DG, Breaux TD (2012) Reconciling multi-jurisdictional legal requirements: A case study in requirements water marking. In: Proceedings of the 20th IEEE International Requirements Engineering Conference (RE'12), pp 91–100

Grossi D, Meyer JJC, Dignum F (2008) The many faces of counts-as: A formal analysis of constitutive rules. Journal of Applied Logic 6(2):192 – 217, DOI https://doi.org/10.1016/j.jal.2007.06.008, URL http://www.sciencedirect.com/science/article/pii/S1570868307000559, selected papers from the 8th International Workshop on Deontic Logic in Computer Science

Hirschberg J, Manning CD (2015) Advances in natural language processing. Science 349(6245):261–266

Hoekstra R, Breuker J, Bello MD, Boer A (2007) The LKIF core ontology of basic legal concepts. In: Proceedings of the 2nd Workshop on Legal Ontologies and Artificial Intelligence Techniques (LOAIT'07), pp 43–63

Hohfeld WN (1917) Fundamental legal conceptions as applied in judicial reasoning. The Yale Law Journal 26(8):710–770

Horty JF (2001) Agency and Deontic Logic. Oxford scholarship online, Oxford University Press, USA

Ingolfo S, Jureta I, Siena A, Perini A, Susi A (2014) Nòmos 3: Legal compliance of roles and requirements. In: Proceedings of the 33rd International Conference on Conceptual Modeling (ER'14), pp 275–288

Jurafsky D, Martin JH (2000) Speech and Language Processing: An Introduction to Natural Language Processing, Computational Linguistics, and Speech Recognition, 1st edn. Prentice Hall PTR, Upper Saddle River, NJ, USA

Kiyavitskaya N, Zeni N, Mich L, Cordy JR, Mylopoulos J (2006) Text mining through semi automatic semantic annotation. In: Proceedings of the 6th International Conference on Practical Aspects of Knowledge Management (PAKM'06), pp 143–154

Kiyavitskaya N, Zeni N, Breaux TD, Antón AI, Cordy JR, Mich L, Mylopoulos J (2008) Automating the extraction of rights and obligations for regulatory compliance. In: Proceedings of the 27th International Conference on Conceptual Modeling (ER'08), pp 154–168

Kummerfeld JK, Hall DLW, Curran JR, Klein D (2012) Parser showdown at the wall street corral: An empirical investigation of error types in parser output. In: Proceedings of the Joint Conference on Empirical Methods in Natural Language Processing and Computational Natural Language Learning (EMNLP-CoNLL'12), pp 1048–1059

Lam H, Hashmi M, Scofield B (2016) Enabling reasoning with LegalRuleML. In: Proceedings of the 10th International Symposium on Rule Technologies. Research, Tools, and Applications (RuleML'16), pp 241–257

Landis J, Koch GG (1977) The measurement of observer agreement for categorical data. Biometrics 33(1):159–174

Levy R, Andrew G (2006) Tregex and tsurgeon: tools for querying and manipulating tree data structures. In: Proceedings of the 5th International Conference on Language Resources and Evaluation (LREC'06), pp 2231–2234

Semantic Legal Metadata Extraction 37Lucassen G, Robeer M, Dalpiaz F, van der Werf JMEM, Brinkkemper S (2017) Extracting conceptual models from user stories with visual narrator. Requirements Engineering 22(3):339–358

Massey A (2012) Legal requirements metrics for compliance analysis. PhD thesis, North Carolina State University, Raleigh, North Carolina, USA

Massey AK, Otto PN, Hayward LJ, Antón AI (2010) Evaluating existing security and privacy requirements for legal compliance. Requirements Engineering 15(1):119–137

Maxwell JC, Antón AI (2010) The production rule framework: developing a canonical set of software requirements for compliance with law. In: Proceedings of the ACM International Health Informatics Symposium (IHI'10), pp 629–636

Maxwell JC, Antón AI, Swire PP, Riaz M, McCraw CM (2012) A legal cross-references taxonomy for reasoning about compliance requirements. Requirements Engineering 17(2):99–115

McDonald RT, Nivre J (2007) Characterizing the errors of data-driven dependency parsing models. In: Proceedings of the 2007 Joint Conference on Empirical Methods in Natural Language Processing and Computational Natural Language Learning (EMNLP-CoNLL'07), pp 122–131

Nivre J, Hall J, Nilsson J, Chanev A, Eryigit G, Kübler S, Marinov S, Marsi E (2007) Maltparser: A language-independent system for data-driven dependency parsing. Natural Language Engineering 13(2):95–135

Peters W, Sagri M, Tiscornia D (2007) The structuring of legal knowledge in LOIS. Artificial Intelligence and Law 15(2):117–135

Petrov S, Barrett L, Thibaux R, Klein D (2006) Learning accurate, compact, and interpretable tree annotation. In: Proceedings of the 21st International Conference on Computational Linguistics and 44th Annual Meeting of the Association for Computational Linguistics (ACL'06)

Pradet Q, Danlos L, de Chalendar G (2014) Adapting verbnet to french using existing resources. In: the Ninth International Conference on Language Resources and Evaluation (LREC'14), pp 1122–1126

Princeton University (2010) About WordNet. http://wordnet.princeton.edu

Quirchmayr T, Paech B, Kohl R, Karey H, Kasdepke G (2018) Semi-automatic rule-based domain terminology and software feature-relevant information extraction from natural language user manuals. Empirical Software Engineering

Rosadini B, Ferrari A, Gori G, Fantechi A, Gnesi S, Trotta I, Bacherini S (2017) Using NLP to detect requirements defects: An industrial experience in the railway domain. In: Proceedings of the 23rd International Working Conference on Requirements Engineering: Foundation for Software Quality (REFSQ'17), pp 344–360

RuleML (2015) Specification of RuleML 1.02. http://wiki.ruleml.org/index.php/Specification_of_RuleML_1.02/

Sagot B (2010) The Lefff, a Freely Available and Large-coverage Morphological and Syntactic Lexicon for French. In: Proceedings of the International Conference on Language Resources and Evaluation(LREC'10), pp 2745–2751

Saldaña J (2015) The Coding Manual for Qualitative Researchers. Sage

Sannier N, Adedjouma M, Sabetzadeh M, Briand LC (2016) Automated classification of legal cross references based on semantic intent. In: Proceedings of the 22nd International Working Conference on Requirements Engineering: Foundation for Software Quality (REFSQ'16), pp 119–134

Sannier N, Adedjouma M, Sabetzadeh M, Briand LC (2017) An automated framework for detection and resolution of cross references in legal texts. Requirements Engineering

22(2):215–237

Sartor G, Casanovas P, Biasiotti M, Fernndez-Barrera M (2013) Approaches to Legal Ontologies: Theories, Domains, Methodologies. Springer

Siena A, Mylopoulos J, Perini A, Susi A (2009) Designing law-compliant software requirements. In: Proceedings of the 28th International Conference on Conceptual Modeling (ER'09), pp 472–486

Siena A, Jureta I, Ingolfo S, Susi A, Perini A, Mylopoulos J (2012) Capturing variability of law with nómos 2. In: Proceedings of the 31st International Conference on Conceptual Modeling (ER'12), pp 383–396

Sleimi A, Sannier N, Sabetzadeh M, Briand LC, Dann J (2018) Automated extraction of semantic legal metadata using natural language processing. In: Proceedings of the 26th IEEE International Requirements Engineering Conference (RE'18), pp 302–311

Sleimi A, Ceci M, Sannier N, Sabetzadeh M, Briand LC, Dann J (2019) A query system for extracting requirements-related information from legal texts. In: Proceedings of the 27th IEEE International Requirements Engineering Conference (RE'19)

Thornton C, Hutter F, Hoos HH, Leyton-Brown K (2013) Auto-weka: Combined selection and hyperparameter optimization of classification algorithms. In: The 19th ACM SIGKDD International Conference on Knowledge Discovery and Data Mining KDD, pp 847–855

Wiki (2004) Wiktionnaire. https://fr.wiktionary.org/

Zeni N, Kiyavitskaya N, Mich L, Cordy JR, Mylopoulos J (2015) GaiusT: supporting the extraction of rights and obligations for regulatory compliance. Requirements Engineering 20(1):1–22

Zeni N, Seid EA, Engiel P, Ingolfo S, Mylopoulos J (2016) Building large models of law with NómosT. In: Proceedings of the 35th International Conference on Conceptual Modeling (ER'16), pp 233–247

## A Appendix: List of modal verbs expressing obligation, permission and prohibition

- sont soumis
- seront soumis
- sera soumis
- est soumis
- est soumise
- sont soumises
- il doit
- elle doit
- ils doivent
- elles doivent
- il devra
- elle devra
- ils devront
- elles devront
- il oblige
- elle oblige
- ils obligent
- elles obligent
- il obligera
- elle obligera
- ils obligeront
- elles obligeront
- obligent
- soumis
- soumise
- doit
- devra
- doivent
- devront
- obligation
- oblig
- devoir
- est tenue
- est tenu
- il est oblig
- elle est oblige
- est oblige
- oblig
- il est ncessaire
- ncessaire
- toujours
- exigence
- astreint
- astreinte
- astreint
- vassur
- assure
- oblige